\documentclass[10pt]{article}
\usepackage{cite} 
\usepackage{graphicx,amsmath,amssymb,amsbsy,latexsym,amsthm}
\usepackage{bm}
\usepackage[mathscr]{eucal}

\theoremstyle{plain}
\newtheorem{definition}{Definition}
\newtheorem{theorem}{Theorem}

\newtheorem{lemma}{Lemma}
\newtheorem{corollary}[theorem]{Corollary}
\newcommand{\startproof}{\noindent\textbf{Proof.} }
\newcommand{\finishproof}{\hfill $\blacksquare$ \\}

\numberwithin{equation}{section}
\def\R{\mathbb{R}}
\def\Z{\mathbb{Z}}
\def\C{\mathbb{C}}

\def\Re{\mathrm{Re}\,}

\def\tr{\mathrm{tr}}
\def\half{\frac{1}{2}}
\def\sgn{\mathrm{sgn}}

\newcommand{\so}{\mathfrak{so}}

\newcommand{\su}{\mathfrak{su}}

\newcommand{\EPRL}{\mathrm{EPRL}}
\newcommand{\prop}{\mathrm{(+)}}
\newcommand{\spin}{\mathfrak{spin}}

\newcommand{\eqa}{\begin{eqnarray}}
\newcommand{\neqa}{\end{eqnarray}}
\newcommand{\be}{\begin{equation}}
\newcommand{\ee}{\end{equation}}

\newcommand{\scr}[1]{\mathcal{#1}}

\newcommand{\dual}{\,\,{}^\star\!}

\newcommand{\dif}{\mathrm{d}}

\newcommand{\Hil}{\mathcal{H}}

\newcommand{\Regge}{\mathrm{Regge}}

\newcommand{\geom}{\mathrm{geom}}
\newcommand{\phys}{\mathrm{phys}}

\newcommand{\sut}[1]{#1}
\newcommand{\cldataSUtwo}{\dummy reduced boundary data \dummy}

\newcommand{\spinsp}{V}
\newcommand{\rotk}{k}  
\newcommand{\simpj}{s}
\newcommand{\genj}{j}
\newcommand{\genjpm}{j}
\newcommand{\lcd}{\eta}
\newcommand{\simp}{S}
\newcommand{\geosimp}{\sigma}
\newcommand{\njm}[3]{|#1; #2,#3\rangle}
\newcommand{\njmb}[3]{\langle #1; #2,#3|}

\newcommand{\ggk}{Y}
\newcommand{\rotgen}{L}
\newcommand{\spinset}{\mathcal{K}_\gamma}
\newcommand{\xmxp}[2]{X^-_{#1#2} X^+_{#2#1}}
\newcommand{\simpiota}[1]{\iota_{\rotk_{#1}}^{\simpj^-_{#1} \simpj^+_{#1}}}
\newcommand{\geniota}[2]{\iota_{\rotk_{#1}}^{\genjpm^-_{#2}, \genjpm^+_{#2}}}
\newcommand{\dummy}{\rule{0mm}{0mm}}

\newcommand{\topdec}[3]{\overset{\scriptscriptstyle #2}{\displaystyle #1}\rule{0pt}{#3}}

\newcommand{\Xdec}[1]{\topdec{X}{#1}{8pt}}
\newcommand{\Gdec}[1]{\topdec{G}{#1}{8pt}}

\newcommand{\genprop}[1]{\, {}^{\{\Lambda_{#1}\}}\!\! A^{\prop}_v}
\newcommand{\genproj}[2]{\, {}^{\{\Lambda_{#1}\}}\! P_{#2}}
\newcommand{\genbeta}[2]{\, {}^{\{\Lambda_{#1}\}}\! \beta_{#2}}

\begin{document}

\title{A proposed proper EPRL vertex amplitude}

\author{Jonathan Engle\thanks{jonathan.engle@fau.edu}
 \\[1mm]
\normalsize \em Department of Physics, Florida Atlantic University, Boca Raton, Florida, 33431 USA}
\date{\today}
\maketitle\vspace{-7mm}

\begin{abstract}

As established in a prior work of the author,
the linear simplicity constraints used in the construction of the so-called `new' spin-foam models
mix three of the five sectors of Plebanski theory as well as two dynamical
orientations, and this is the reason for multiple terms
 in the asymptotics of the EPRL vertex amplitude as calculated by Barrett et al.
Specifically, the term equal to the usual exponential of $i$ times the Regge action corresponds to configurations either in sector (II+) with positive orientation or sector (II-) with negative orientation.  The presence of the other terms beyond this
cause problems in the semiclassical limit of the spin-foam model when considering multiple 4-simplices due to the
fact that the different terms for different 4-simplices mix in the semiclassical limit,
leading in general to a non-Regge action and hence non-Regge and nongravitational configurations
persisting in the semiclassical limit.

To correct this problem, we propose to modify the vertex so its asymptotics include only the one term of the form
$e^{iS_{\Regge}}$.
To do this, an explicit classical discrete condition is derived that isolates the desired gravitational
sector corresponding to this one term.  This condition is quantized
and used to modify the vertex amplitude, yielding what we call the `proper EPRL vertex amplitude.'
This vertex still depends only on standard $SU(2)$ spin-network data on the boundary,
is $SU(2)$ gauge-invariant, and is linear in the boundary state, as required. In addition, the asymptotics now consist
in the single desired term of the form $e^{iS_{\Regge}}$, and all degenerate configurations are
exponentially suppressed.  A natural generalization to the Lorentzian signature is also presented.
\end{abstract}


\section{Introduction}

At the heart of the path integral formulation of quantum mechanics \cite{feynman1948, feynman1942}
is the prescription that the contribution to the transition amplitude by each classical trajectory should be the exponential of $i$ times the classical action. The use of such an expression has roots tracing back to Paul Dirac's \textit{Principles of Quantum Mechanics} \cite{dirac1930}, and
is central to the successful derivation of the classical limit of the path integral, using the fact that
the classical equations of motion are the stationary points of the classical action.

The modern spin-foam program  \cite{rovelli2004, perez2003, rovelli2011}
aims to provide a definition, via path integral, of the dynamics of loop quantum gravity (LQG)
\cite{al2004, rovelli2004, thiemann2007, rovelli2011},
a background independent canonical quantization of general relativity.
The only spin-foam model to so far match the kinematics of loop quantum gravity and therefore achieve this
goal is the so-called EPRL model \cite{elpr2007, epr2007, epr2007a, kkl2009}, which,
for Barbero-Immirzi parameter less than 1 is equal to the FK model \cite{fk2007}.

In loop quantum gravity, geometric operators have \textit{discrete} spectra. The basis of states diagonalizing the area and other geometric operators are the \textit{spin-network states}.  The spin-foam path integral consists in a sum over amplitudes associated to histories of such states, called \textit{spin-foams}. Each spin-foam in turn can be interpreted in terms of a Regge geometry on
a simplicial lattice.  The simplest amplitude provided by a spin-foam model is the so-called \textit{vertex amplitude} which gives the probability amplitude for a set of quantum data on the boundary of single 4-simplex.

The semiclassical (i.e. large quantum number, equivalent to $\hbar \rightarrow 0$) limit \cite{bdfgh2009} of the EPRL vertex amplitude, however, is \textit{not} equal to the exponential of $i$ times the
Regge action as one would desire, but includes other terms as well.\footnote{From 
\cite{cf2008}, this is true also for the Freidel-Krasnov model, as must be the case as it is
equal to EPRL for $\gamma < 1$. In \cite{cf2008}, one finds two terms, not one, in the asymptotics.
Furthermore, the presence of only two terms is likely due to their reformulating the model as a discrete first order path integral and then
imposing nondegeneracy, a procedure whose equivalent in the spin-foam language, needed for contact with canonical states,
is not known \cite{cf2008}.}
As a consequence, when considering multiple 4-simplices, the semiclassical limit of the amplitude has cross-terms, each of which consists in the exponential
of a sum of terms, one for each 4-simplex, equal to the Regge action for that 4-simplex times differing coefficients, yielding what can be called a
`generalized Regge action'\cite{mp2011, hz2011}.
The stationary point equations of this `generalized Regge action' are \textit{not} the Regge equations of motion and hence \textit{not} those of
general relativity, whence general relativity will fail to be recovered in the classical limit.
As presented in the recent work \cite{engle2011, engle2011err}, the extra terms causing this problem correspond to different sectors of Plebanski theory, as well as different orientations of the space-time.  These various sectors and orientations are present in the spin-foam sum
because the so-called \textit{linear simplicity constraint} --- the constraint which is also used in the
Freidel-Krasnov model \cite{fk2007} --- allows them.

In this paper, we propose a modification to the EPRL vertex amplitude which solves this problem.
We begin by deriving, at the classical discrete level, a condition which isolates the sector corresponding only to
the first term in the asymptotics, the exponential of $i$ times the Regge action. We call this sector the `Einstein-Hilbert'
sector, because it is the sector of Plebanski theory in which the BF action reduces to the Einstein-Hilbert action.
More specifically, this sector consists in configurations which are either in (what is called) Plebanski sector (II+) with positive space-time orientation, or (what is called) Plebanski sector (II-) with negative orientation.\footnote{In a prior
version of this article, the sector corresponding to the first term in the asymptotics was 
mischaracterized as the (II+) sector, whereas in fact it is the combination of sectors stated here.  
This mistake was due to an error in the prior work \cite{engle2011} which was corrected in \cite{engle2011err}.
The correction of this error did not at all change the proper vertex or its motivation rooted in the semiclassical 
limit, but only changed the interpretation in terms of Plebanski sectors and orientations.
}
This condition is then appropriately quantized and inserted into the expression for the vertex, leading to a modification of the EPRL vertex amplitude.
The resulting vertex continues to be a function of a loop quantum gravity boundary state and hence may still be used to define dynamics \textit{for loop quantum gravity}.
It furthermore remains
linear in the boundary state and fully $SU(2)$ invariant --- two conditions forming a nontrivial
%
%
requirement
restricting the possible expressions for the vertex.
It is also in a 
precise sense $Spin(4)$ invariant.
Lastly, as is shown in the final section of this paper, for a complete set of boundary states, the
asymptotics of the vertex include only a single term, equal to
the exponential of $i$ times the Regge action, enabling the correct equations of motion to dominate
in the classical limit.
We call the resulting vertex amplitude the \textit{proper EPRL vertex amplitude}.
A natural generalization to the Lorentzian case is presented in section \ref{lorsect}.
A summary of these results can be found in \cite{engle2012}.

We begin the paper with a review of the classical discrete framework underlying the spin-foam model and derive the
condition isolating the Einstein-Hilbert sector. Then, after briefly reviewing the existing EPRL vertex
amplitude, the definition of the new proper vertex is introduced. The last half of the paper
is then spent proving the properties summarized above.  We then close with a discussion.

\section{Classical analysis}
\label{clsect}

\subsection{Background}

\subsubsection{Generalities}

We use the same definitions as in \cite{engle2011}.
%
%
Let $\sut{\tau}^i := \frac{-i}{2}\sut{\sigma}^i$ ($i = 1,2,3$),
where $\sut{\sigma}^i$ are the Pauli matrices.
For each element $\sut{\lambda} \in \su(2)$, $\lambda^i \in \R^3$
shall denote its components with respect to the basis $\sut{\tau}^i$.
Let $I$ denote the $2\times 2$ identity matrix.
%
%
We also freely use the
isomorphism between $\spin(4):= \su(2) \oplus \su(2)$ and $\so(4)$,
$(\sut{J}_-,\sut{J}_+) \equiv (J_-^i\sut{\tau}_i, J_+^i\sut{\tau}_i) \leftrightarrow J^{IJ}$
($I,J  = 0,1,2,3$), explicitly given by
\begin{eqnarray}
\nonumber
J^{ij} &=& \epsilon^{ij}{}_k (J_+^k + J_-^k) \\
\label{alginv}
J^{0i} &=& J_+^i - J_-^i .
\end{eqnarray}
$J_+^i$ and $J_-^i$ are called the self-dual and anti-self-dual parts of $J^{IJ}$.
Furthermore, we remind the reader \cite{bdfgh2009}
of the explicit expression for
the action of $Spin(4) = SU(2) \times SU(2)$ group elements on $\R^4$. For each $V^I \in \R^4$ define
\begin{equation}
\label{zetadef}
\zeta(V):= V^0 I + i \sigma_i V^i .
\end{equation}
Then the action of $G = (X^-, X^+)$ is given by
\begin{equation}
\label{sofouract}
\zeta(G \cdot V) = X^- \zeta(V) (X^+)^{-1}.
\end{equation}
%
%

\subsubsection{Discrete classical framework}
\label{clframe}

Spin-foam models of quantum gravity are based on a formulation of
gravity as a \textit{constrained BF theory}, using the ideas of Plebanski \cite{plebanski1977}.
In the continuum, the basic variables are an $\so(4)$ connection $\omega_\mu^{IJ}$ and an $\so(4)$-valued two-form $B_{\mu\nu}^{IJ}$, which we call the \textit{Plebanski two-form}, where lower case greek letters are used for space-time manifold
indices.  The action is
\begin{equation}
\label{BFaction}
S = \frac{1}{2\kappa}\int (B + \frac{1}{\gamma} \dual B)_{IJ} \wedge F^{IJ},
\end{equation}
with $F := \dif \omega + \omega \wedge \omega$ the curvature of $\omega$,
$\dual$ the Hodge dual on internal indices $I,J,K \dots$,
$\kappa := 8\pi G$, and $\gamma \in \R^+$ the Barbero-Immirzi parameter.
If $B_{\mu\nu}^{IJ}$ satisfies what we call the Plebanski constraint \cite{df1998, bhnr2004},
 it must be one of the five forms
\begin{description}
\item[(I$\pm$)] $B^{IJ} = \pm e^I \wedge e^J$ for some $e^I_\mu$
\item[(II$\pm$)] $B^{IJ} = \pm \frac{1}{2} \epsilon^{IJ}{}_{KL} e^K \wedge e^L$
for some $e^I_\mu$
\item[(deg)] $\epsilon_{IJKL}\lcd^{\mu\nu\rho\sigma} B^{IJ}_{\mu\nu} B^{KL}_{\rho\sigma} = 0$ (degenerate case)
%
%
\end{description}
which we call  \textit{Plebanski sectors}.
Here $\epsilon_{IJKL}$ denotes the internal Levi-Civita array, and $\lcd^{\mu\nu\rho\sigma}$
denotes the Levi-Civita tensor of density weight $1$.
In sectors (II$\pm$), the BF action reduces to
%
%
%
a sign times the \textit{Holst action} for gravity \cite{holst1995},
\begin{equation}
S_{Holst} = \frac{1}{4\kappa} \int \left(\epsilon_{IJKL} e^K \wedge e^L + \frac{2}{\gamma} e_I \wedge e_J\right)
\wedge F^{IJ},
\end{equation}
the Legendre transform of
which forms the starting point for loop quantum gravity \cite{holst1995, al2004}.

In spin-foam quantization, one usually introduces a simplicial discretization of space-time.
However, in this paper we concern ourselves with the so-called `vertex amplitude',
which may be thought of as the transition amplitude for a single 4-simplex. For clarity,
we thus focus on a single oriented 4-simplex $\simp$.  The EPRL model
has also been generalized to general cell-complexes \cite{kkl2009}; however because we use the
work \cite{bdfgh2009}, and because we introduce formulae that, so far, apply only to
4-simplices, we restrict the discussion to the case of a 4-simplex.
In $\simp$, number the tetrahedra
%
%
$a=0, \dots ,4$,\footnote{
In the prior work \cite{engle2011}, the order of the numbering was used to code the orientation of $S$.
This was done by imposing an \textit{ordering condition} correlating the orientation of $S$ to the numbering.
However, in this paper, we present things in such a way that one does not need to code
the orientation in the numbering, and so the numbering
is left arbitrary.}
%
%
and let $\Delta_{ab}$ denote the triangle between tetrahedra $a$ and $b$, oriented as part of the
boundary of $a$.
One thinks of each tetrahedron, as well as the 4-simplex itself, as having its own `frame' \cite{epr2007}.
One has a parallel transport map from each tetrahedron to the 4-simplex
frame, yielding in our case 5 parallel transport maps $G_a = (X_a^-,X_a^+) \in Spin(4)$, $a = 0, \dots ,4$.
The continuum two-form $B$ is then represented by the algebra elements
$B_{ab} = \int_{\Delta_{ab}} B$,
where each element is treated as being `in the frame at $a$.'
For each $ab$, in terms of self-dual and anti-self-dual parts, these elements are related
to the momenta conjugate to the parallel transports (see section \ref{cansect}) by
\cite{engle2011, elpr2007}
\begin{equation}
\label{sdasd_rel}
(J^\pm_{ab})^i = \left(\frac{\gamma \pm 1}{\kappa\gamma}\right) (B^{\pm}_{ab})^i.
\end{equation}
We call $B_{ab}$ and $J_{ab}$ the \textit{canonical bivectors} due to their
role in the canonical theory in section \ref{cansect}.

From the discrete data $\{B_{ab}^{IJ}, G_a\}$
one can reconstruct the continuum two-form $B_{\mu\nu}^{IJ}$ as follows. Fix
a flat connection $\partial_\mu$ on the 4-simplex $\simp$, such that
$\simp$ is the convex hull of its vertices as determined by the affine structure defined by $\partial_\mu$; we say such a flat connection is \textit{adapted} to $S$.
The choice of such a connection is unique up to diffeomorphism and hence is a
pure gauge choice (see appendix \ref{defapp}).
If the data $\{B_{ab}^{IJ}, G_a\}$ satisfy (1.) closure, $\sum_{b \neq a} B_{ab}^{IJ}=0$, and (2.) orientation,
$G_a \triangleright B_{ab} = - G_b \triangleright B_{ba}$, then
it has been proven \cite{bfh2009, engle2011} that
%
%
%
there exists a unique two-form field $B_{\mu\nu}^{IJ}$ on the manifold $\simp$,
constant with respect to $\partial_\mu$, such that
\begin{equation}
\label{Brecon}
\mathcal{B}_{ab} := G_a \triangleright B_{ab} = \int_{\Delta_{ab}} B
\end{equation}
for all $a \neq b$.
Here the left hand side is the  parallel transport of the bivectors $B_{ab}^{IJ}$
to the `4-simplex frame', henceforth denoted $\mathcal{B}_{ab}$,
and $\triangleright$ here and throughout the rest of the paper denotes the adjoint action.
Both closure (1.) and orientation (2.) are imposed in the EPRL vertex
in the sense that violations are suppressed exponentially \cite{bdfgh2009}.
In addition, the EPRL model  imposes (3.) \textit{linear simplicity},
\begin{equation}
\label{gfconstr}
C_{ab}^I := \half \scr{N}_J \epsilon^{JI}{}_{KL} B_{ab}^{KL} \approx 0 ,
\end{equation}
where $\scr{N}^I:= (1,0,0,0)$,
as a restriction on the allowed boundary states for each 4-simplex, as shall be reviewed in the quantum theory below.
From (\ref{gfconstr}), it follows that the continuum two-form $B_{\mu\nu}^{IJ}$ defined by (\ref{Brecon}) is
in Plebanski sector (II+), (II-) or (deg) \cite{engle2011}.  We represent this sector by a function $\nu(B_{\mu\nu})$, defined to be $+1$ if $B_{\mu\nu}$ is in (II+), $-1$
if $B_{\mu\nu}$ is in (II-), and $0$ if $B_{\mu\nu}$ is degenerate.
If $\nu(B_{\mu\nu}) \neq 0$, $B_{\mu\nu}$ furthermore defines an
\textit{orientation} of $S$,
which can either agree or disagree with the fixed orientation of $S$ used to define form integrals.
We represent this dynamically defined orientation by its sign relative to the fixed orientation $\mathring{\epsilon}^{\mu\nu\rho\sigma}$ of $S$:
\begin{equation}
\label{ordef}
\omega(B_{\mu\nu}) := \sgn (\mathring{\epsilon}^{\mu\nu\rho\sigma} \epsilon_{IJKL} B_{\mu\nu}^{IJ} B_{\rho\sigma}^{KL}),
\end{equation}
where, for convenience, $\sgn(\cdot)$ is defined to be zero when its argument is zero.
%
%
Because the only arbitrary choice in the construction of $B_{\mu\nu}^{IJ}$, that of the flat connection $\partial_\mu$, 
is unique up to
diffeomorphism, a diffeomorphism which, when chosen to preserve each face of $S$, must be orientation preserving, and because each Plebanski sector as well as the dynamically determined orientation is invariant under such diffeomorphisms, the
functions $\nu(B_{\mu\nu}(\{\mathcal{B}_{ab}\}, \partial))$  and $\omega(B_{\mu\nu}(\{\mathcal{B}_{ab}\}, \partial))$ are independent of the choice of connection
$\partial_\mu$ adapted to $S$, so that one can write simply
$\nu(\{\mathcal{B}_{ab}\})$ and $\omega(\{\mathcal{B}_{ab}\})$.
(For a more detailed derivation of this fact, see appendix \ref{defapp}.)
%
%
This reviews the sense, established in \cite{engle2011},
in which the \textit{classical} constraints
imposed \textit{quantum} mechanically in the EPRL model admit the three distinct, well-defined Plebanski sectors (II+), (II-), and (deg), as well as two possible dynamical orientations.\footnote{In a different sense based on discrete analogies, an awareness of the presence of the three Plebanski sectors was implicit already in \cite{cf2008}.}

\subsubsection{Reduced boundary data}

The set of canonical bivectors $B_{ab}^{IJ}$ satisfying linear simplicity is parameterized by
what we call \textit{reduced boundary data} ---
one unit 3-vector $n^i_{ab}$ per ordered pair $ab$, and one area $A_{ab}$ per triangle
$(ab)$ --- via
\begin{equation}
\label{Bparam}
B_{ab} = \frac{1}{2}A_{ab}(-n_{ab}, n_{ab}) .
\end{equation}
From (\ref{sdasd_rel}) and (\ref{Bparam}), the generators of internal spatial rotations in terms of the
\cldataSUtwo are
\begin{equation}
\label{classLdef}
L^i_{ab} = (J^-)^i_{ab} + (J^+)^i_{ab} = \frac{1}{\kappa \gamma} A_{ab} n^i_{ab}.
\end{equation}
The corresponding bivectors
in the 4-simplex frame then take the form
\begin{equation}
\label{physbiv}
\mathcal{B}_{ab} =  B_{ab}^\phys(A_{ab},\sut{n}_{ab},G_a):=\frac{1}{2}A_{ab}(-X^-_a \triangleright n_{ab}, X^+_a \triangleright n_{ab}) .
\end{equation}
We call (\ref{physbiv}) the `physical' bivectors reconstructed from $A_{ab},\sut{n}_{ab},G_a$.
In terms of the reduced boundary data, closure and orientation become the conditions $\sum_{b \neq a} A_{ab} n_{ab} =0$
and $X^\pm_{a} \triangleright n_{ab} = - X^{\pm}_{b} \triangleright n_{ba}$.

\subsubsection{Reconstruction theorem}

%
%
In addition to reconstructing the 2-form field $B_{\mu\nu}^{IJ}$
from the bivectors $\mathcal{B}_{ab} = B_{ab}^\phys(A_{ab}, n_{ab}, G_a)$,
one can also reconstruct a geometrical 4-simplex in $\R^4$.
This will be needed in the present paper.
Let $M$ denote $\R^4$ as an oriented manifold, equipped with the
canonical $\R^4$ metric.
A \textit{geometrical 4-simplex} $\sigma$ in $M$ is the convex hull of 5 points, called vertices,
in $M$, not all of which lie in the same 3-plane.
We define a \textit{numbered 4-simplex} $\sigma$ to be a geometrical 4-simplex with tetrahedra
numbered $0, \dots 4$.
Given a numbered 4-simplex  in $M$, the associated
\textit{geometrical bivectors} $(B_{ab}^{\geom})^{IJ}$
are defined as
$(B_{ab}^{\geom})^{IJ}:= A(\Delta_{ab}) \frac{(N_a \wedge N_b)^{IJ}}{|N_a \wedge N_b|}$,
where $A(\Delta_{ab})$ is the area of the triangle $\Delta_{ab}$
shared by tetrahedra $a$ and $b$, and
$N_a^I$ is the outward unit normal to tetrahedron $a$,
$(N_a \wedge N_b)^{IJ} := 2N_a^{[I} N_b^{J]}$, and
$|X^{IJ}|^2 := \half X^{IJ}X_{IJ}$.

A set of reduced boundary data
$\{A_{ab}, \sut{n}_{ab}\}$  is \textit{nondegenerate} if,
for each $a$, the span of the vectors $\sut{n}_{ab}$ with $b\neq a$
is three dimensional. We call two sets of $SU(2)$ group elements $\{U^1_a\}, \{U^2_a\}$
\textit{equivalent}, $\{U^1_a\} \sim \{U^2_a\}$, if
$\exists \ggk \in SU(2)$ and five signs $\epsilon_a$ such that
\begin{equation}
\label{eqrel}
U^2_a = \epsilon_a \ggk U^1_a.
\end{equation}
For the proof of the following partial version of theorem 3 in \cite{bdfgh2009}, see
\cite{bdfgh2009, engle2011}.
\begin{theorem}[Partial version of the reconstruction theorem]
\label{reconth}
Let a set of nondegenerate reduced boundary data
$\{A_{ab},\sut{n}_{ab}\}$ satisfying closure be given, as well as
a set $\{G_a\} \subset Spin(4)$, $a=0,\dots,4$, solving the
orientation constraint, such that
$\{X_a^-\} \not\sim \{X_a^+\}$.  Then there exists a numbered 4-simplex $\geosimp$ in $\R^4$, unique up to
inversion and translation,
such that
\begin{equation}
\label{reconpart}
B_{ab}^{\phys}(A_{ab},\sut{n}_{ab},G_a) = \mu B_{ab}^{\geom}(\geosimp)
\end{equation}
for some $\mu = \pm 1$, with $\mu$ independent of the ambiguity in $\geosimp$.
\end{theorem}
The sign $\mu$ in the above theorem is uniquely determined by the data
$\{A_{ab}, n_{ab}, G_a\}$.  In fact, as shown in \cite{engle2011err},
it is equal to the \textit{product} of the sign corresponding to
the Plebanski sector $\nu(B^\phys_{ab}(A_{ab}, n_{ab}, G_a))$ and the sign of the orientation
$\omega(B^\phys_{ab}(A_{ab}, n_{ab}, G_a))$.
Recall we have defined the Einstein-Hilbert sector to include two-forms 
$B_{\mu\nu}$ which are either in Plebanski sector (II+) with
positive orientation or in Plebanski sector (II-) with negative orientation.
The continuum two form $B_{\mu\nu}$ reconstructed from the bivectors
$\{B^\phys_{ab}(A_{ab}, n_{ab}, G_a)\}$
will thus be in the Einstein-Hilbert sector (in which case we also say the bivectors
are in the Einstein-Hilbert sector) \textit{if and only if}
$\mu = \nu \omega = +1$.

\subsection{Explicit classical expression for the geometrical bivectors, and the restriction to the Einstein-Hilbert sector}

We now come to the new part of the classical analysis.

\begin{lemma}
\label{normlem}
Let $\{A_{ab}, n_{ab}, G_a\}$ be given satisfying the hypotheses of theorem \ref{reconth} and let $\sigma$
be the numbered 4-simplex guaranteed to exist by this theorem.  Let $\{N_a^I\}$ denote the outward pointing
normals to the tetrahedra of $\sigma$.  Then
\begin{equation}
N^I_a = \alpha_a (G_a \cdot \scr{N})^I
\end{equation}
for some set of signs $\alpha_a$.
\end{lemma}
{\startproof
We first note that
\begin{eqnarray*}
(N_a \wedge N_b)^{IJ} (G_a \cdot \scr{N})_J &\propto& B^\phys_{ab}(A_{ab}, n_{ab}, X_{ab}^\pm)^{IJ} (G_a \cdot \scr{N})_J \\
&\propto& \left[G_a \triangleright (-n_{ab}, n_{ab}) \right]^{IJ} (G_a \cdot \scr{N})_J \\
&=& (G_a)^I{}_K (G_a)^J{}_L (-n_{ab}, n_{ab})^{KL} (G_a)_{IM} \scr{N}^M \\
&=& (G_a)^J{}_L (-n_{ab},n_{ab})^{KL} \scr{N}_K = (G_a)^J{}_L (-n_{ab}, n_{ab})^{0L} = 0
\end{eqnarray*}
where (\ref{reconpart}) was used in the first line, and (\ref{alginv}) was used in the last line.
Since this holds for all $b$, it follows that
$G_a \cdot \scr{N}$ is proportional to $N_a$; as both of these vectors are unit, the the coefficient of proportionality
must be $\pm 1$ for each $a$.
\finishproof}

\newcommand{\heq}{\widehat{=}}
For the following theorem and throughout the rest of the paper,
let $\heq$ denote equality modulo multiplication by a \textit{positive} real number.
\begin{theorem}
\label{geombivth}
Let $\{A_{ab}, n_{ab}, G_a\}$ be given satisfying the hypotheses of theorem \ref{reconth} and  let $\sigma$
be the numbered 4-simplex guaranteed to exist by this theorem.  Then
\begin{equation}
\label{Bgeomeq}
B^\geom_{ab}(\sigma) \heq \beta_{ab}(\{G_{a'b'}\}) (G_a \cdot \scr{N}) \wedge (G_b \cdot \scr{N})
\end{equation}
where
\begin{eqnarray}
\nonumber
\dummy \hspace{-0.6cm}\beta_{ab}(\{G_{a'b'}\}) \hspace{-0.35cm} &\equiv& \hspace{-0.35cm} \beta_{ba}(\{G_{a'b'}\})\\
\label{betadef}
&:=&\hspace{-0.35cm} -\sgn\left[\epsilon_{ijk} (G_{ac}\cdot\scr{N})^i  (G_{ad}\cdot\scr{N})^j  (G_{ae}\cdot\scr{N})^k
\epsilon_{lmn} (G_{bc}\cdot\scr{N})^l  (G_{bd}\cdot\scr{N})^m  (G_{be}\cdot\scr{N})^n\right]
\end{eqnarray}
with $\{c,d,e\} = \{0, \dots, 4 \} \setminus \{a,b\}$ in any order, and $\sgn$ is defined to be zero when its argument is zero.
\end{theorem}
{\startproof
Let $\{N_a^I\}$ be the outward pointing
normals to the tetrahedra of $\sigma$.  Then they satisfy the four-dimensional
closure relation (see appendix \ref{closureapp})
\begin{equation}
\sum_a V_a N_a^I = 0
\end{equation}
where $V_a > 0$ is the volume of the $a$th tetrahedron, implying
\begin{equation}
N_a^I = - \frac{1}{V_a} \sum_{a' \neq a} V_{a'} N_{a'}^I.
\end{equation}
Thus
\begin{eqnarray*}
0 &<&  \epsilon(N_a, N_c, N_d, N_e)^2
= - \frac{V_b}{V_a}  \epsilon(N_b, N_c, N_d, N_e) \epsilon(N_a, N_c, N_d, N_e) \\
&\heq& -  \alpha_a \alpha_b \epsilon(G_b\cdot \scr{N}, G_c\cdot \scr{N}, G_d\cdot \scr{N}, G_e\cdot \scr{N})
\epsilon(G_a\cdot \scr{N}, G_c\cdot \scr{N}, G_d\cdot \scr{N}, G_e\cdot \scr{N}) \\
&=&  -  \alpha_a \alpha_b \epsilon(\scr{N}, G_{bc}\cdot \scr{N}, G_{bd}\cdot \scr{N}, G_{be}\cdot \scr{N})
 \epsilon(\scr{N}, G_{ac}\cdot \scr{N}, G_{ad}\cdot \scr{N}, G_{ae}\cdot \scr{N}) \\
&=&  -  \alpha_a \alpha_b
 \epsilon_{ijk} (G_{bc}\cdot \scr{N})^i  (G_{bd}\cdot \scr{N})^j  (G_{be}\cdot \scr{N})^k
 \epsilon_{lmn} (G_{ac}\cdot \scr{N})^l  (G_{ad}\cdot \scr{N})^m  (G_{ae}\cdot \scr{N})^n
\end{eqnarray*}
where $\{\alpha_a\}$ are the signs in lemma \ref{normlem}.
Therefore
\begin{equation}
\beta_{ab}(\{G_{a'b'}\}) = \alpha_a \alpha_b
\end{equation}
where $\beta_{ab}(\{G_{a'b'}\})$ is as in (\ref{betadef}). We thus have
\begin{displaymath}
B^\geom_{ab}(\sigma) \heq
N_a \wedge N_b = \alpha_a \alpha_b  (G_a \cdot \scr{N}) \wedge (G_b \cdot \scr{N})
= \beta_{ab}(\{G_{a'b'}\}) (G_a \cdot \scr{N}) \wedge (G_b \cdot \scr{N}).
\end{displaymath}
\finishproof}
\noindent Throughout this paper, let $\beta_{ab}(\{G_{a'b'}\})$
be defined by (\ref{betadef}), and for convenience we define \\
\mbox{$\tilde{B}^\geom_{ab}(G_{a'}) := \beta_{ab}(\{G_{a'b'}\}) (G_a \cdot \scr{N}) \wedge (G_b \cdot \scr{N})$},
the right hand side of (\ref{Bgeomeq}).

Because the expression $(G \cdot \scr{N})^i$ used above will appear often, it is useful to stop for a moment to prove some facts about it.
From (\ref{zetadef}) and (\ref{sofouract}),
\begin{displaymath}
(G_{ab} \scr{N})^0 I + i \sigma_i (G_{ab} \cdot \scr{N})^i = \zeta(G_{ab} \cdot \scr{N}) = X_{ab}^- X_{ba}^+,
\end{displaymath}
from which one obtains the alternate expression
\begin{equation}
(G_{ab} \cdot \scr{N})^i = \tr(\tau^i X_{ab}^- X_{ba}^+) .
\end{equation}
The meaning of this latter expression in turn is made clear in the following definition.
\begin{definition}
\label{propaxis}
Given $g \in SU(2)$ not equal to $\pm I$, there exists a unique unit vector $n[g]^i \in \R^3$ and $\alpha[g] \in (0,2\pi)$ satisfying
\begin{equation}
\label{SUtwoexp}
g = \exp(\alpha[g] \cdot n[g] \cdot \sut{\tau}) = \cos\left(\frac{\alpha[g]}{2}\right)
+ i n[g] \cdot \sut{\sigma} \sin\left(\frac{\alpha[g]}{2}\right) .
\end{equation}
%
%
We call $n[g]^i$ the \textit{proper axis} of $g$.
\end{definition}
\noindent In terms of the above definition, one has
\begin{equation}
\label{expl_axis}
(G_{ab} \cdot \scr{N})^i = \tr(\tau_i \xmxp{a}{b}) = \sin\left(\frac{\alpha[\xmxp{a}{b}]}{2}\right) n[\xmxp{a}{b}]^i .
\end{equation}

\begin{lemma}
\label{mulemm}
Let $\{A_{ab}, n_{ab}, G_a\}$ be given satisfying the hypotheses of
theorem \ref{reconth} and let $\sigma$ be the numbered 4-simplex thereby
guaranteed to exist.
%
%
Then
\begin{equation}
\label{mueq}
\mu = B^\geom_{ab}(\sigma)_{IJ} B^\phys_{ab}(A_{ab}, n_{ab}, G_a)^{IJ} \heq
\beta_{ab}(\{G_{a'b'}\}) \tr(\tau_i X^-_{ab} X^+_{ba}) L^i_{ab} .
\end{equation}
\end{lemma}
{\startproof
Starting from (\ref{reconpart}) and theorem \ref{geombivth},
\begin{eqnarray}
\nonumber
\mu &=& B^\geom_{ab}(\sigma)_{IJ} B^\phys_{ab}(A_{ab}, n_{ab}, G_a)^{IJ} \\
\nonumber
&\heq& \beta_{ab}(\{G_{a'b'}\}) \left[(G_a\cdot \scr{N}) \wedge (G_b \cdot \scr{N})\right]_{IJ}
\half A_{ab} \left[G_a \triangleright (-n_{ab}, n_{ab})\right]^{IJ} \\
\nonumber
&=& \half A_{ab} \beta_{ab}(\{G_{a'b'}\}) \left[\scr{N} \wedge (G_{ab} \cdot \scr{N})\right]_{IJ}
 \left[-n_{ab}, n_{ab}\right]^{IJ} \\
\nonumber
&=& A_{ab} \beta_{ab}(\{G_{a'b'}\}) \left[\scr{N} \wedge (G_{ab} \cdot \scr{N})\right]_{0i}
 \left[-n_{ab}, n_{ab}\right]^{0i} \\
\nonumber
&=& 2 A_{ab} \beta_{ab}(\{G_{a'b'}\}) (G_{ab} \cdot \scr{N})_i n_{ab}^i
\heq \beta_{ab}(\{G_{a'b'}\}) (G_{ab} \cdot \scr{N})_i L_{ab}^i \\
\nonumber
&=&  \beta_{ab}(\{G_{a'b'}\}) \tr(\tau_i X^-_{ab} X^+_{ba})  L_{ab}^i   .
\end{eqnarray}
\finishproof}

\noindent We now come to the classical condition isolating the Einstein-Hilbert sector.
\begin{theorem}
\label{classcon}
Let a set of nondegenerate reduced boundary data
$\{A_{ab}, n_{ab}\}$ satisfying closure be given,
as well as a set $\{G_a\} \subset Spin(4), a = 0, \dots 4$
solving the orientation constraint. Then
$B^\phys_{ab}(A_{ab}, n_{ab}, G_a)$ is in the Einstein-Hilbert sector 
(that is, $\mu = \omega\nu = +1$) iff
\begin{equation}
\label{EHineq}
\beta_{ab}(\{G_{a'b'}\}) \tr(\tau_i X^-_{ab} X^+_{ba}) L^i_{ab} > 0
\end{equation}
for any one pair $a,b$.
\end{theorem}
{\startproof

($\Rightarrow$) Suppose $B_{ab}^\phys(A_{ab}, n_{ab}, G_a)^{IJ}$
is in the Einstein-Hilbert sector.  Then by theorem 3 in \cite{engle2011, engle2011err},
$\{X^-_a\} \not\sim \{X^+_a\}$, so that $\mu$ exists, and $\mu = 1$. Lemma \ref{mulemm}
then
implies (\ref{EHineq}).

($\Leftarrow$) Suppose (\ref{EHineq}) holds.  Suppose by way of contradiction $\{X^-_a\} \sim \{X^+_a\}$.
Then $\tr(\tau^i X^-_{ab} X^+_{ba}) = 0$ contradicting (\ref{EHineq}).  Therefore $\{X^-_a\} \not\sim \{X^+_a\}$.
Lemma \ref{mulemm}
together with (\ref{EHineq}) then implies $\mu = +1$,
so that theorem 3 in \cite{engle2011, engle2011err} implies  $B_{ab}^\phys(A_{ab}, n_{ab}, G_a)^{IJ}$ is in
the Einstein-Hilbert sector.
\finishproof}

\section{Review of quantum framework and the EPRL vertex}

\subsection{Notation for $SU(2)$ and $Spin(4)$ structures.}
\label{notationsec}

Let $V_j$ denote the carrying space for the spin $j$ representation of $SU(2)$,
and $\rho_\genj(g), \rho_\genj(\sut{x})$
the representation of $g \in SU(2)$ and $x \in \su(2)$ thereon,
with the $\genj$ subscript dropped when it is clear from the context.
%
%
Let $\hat{\rotgen}^i:= i \rho(\sut{\tau}^i)$ denote the generators in each of these representation
according to the context.  Let $\epsilon : \spinsp_\genj \times \spinsp_\genj \rightarrow \C$
denote the invariant bilinear epsilon inner
product, and $\langle \cdot , \cdot \rangle$ the Hermitian inner product, on
$\spinsp_\genj$ \cite{rovelli2004, bdfgh2009}.
These inner products determine an
antilinear structure map $J: \spinsp_\genj \rightarrow \spinsp_\genj$ by
$\epsilon(\psi, \phi) = \langle J \psi, \phi \rangle$.
%
%
%
$J$ commutes with all group representation matrices,
so that it anticommutes with all generators.

Let $\spinsp_{\genjpm^-,\genjpm^+} = \spinsp_{\genjpm^-} \otimes \spinsp_{\genjpm^+}$ denote
the carrying space for the spin $(\genjpm^-,\genjpm^+)$ representation of $Spin(4) \equiv SU(2) \times SU(2)$,
and $\rho_{\genjpm^-,\genjpm^+}(X^-,X^+) := \rho_{\genjpm^-}(X^-)\otimes\rho_{\genjpm^+}(X^+)$ the representation of
$(X^-,X^+) \in Spin(4)$ thereon, again with the subscript dropped
when it is clear from the context.
$\hat{J}^i_- := i\rho(\sut{\tau}^i) \otimes I_{\genjpm^+}$ and $\hat{J}^i_+ := i I_{\genjpm^-} \otimes \rho(\sut{\tau}^i)$
are then the anti-self-dual and self-dual generators
respectively, so that
$\hat{\rotgen}^i := \hat{J}^i_- + \hat{J}^i_+$
are the generators of spatial rotations on $\spinsp_{\genjpm^-, \genjpm^+}$.
%
%
Define the bilinear form $\epsilon : \spinsp_{\genjpm^+,\genjpm^-} \times \spinsp_{\genjpm^+,\genjpm^-} \rightarrow \C$ by
$\epsilon(\psi^+ \otimes \psi^-, \phi^+ \otimes \phi^-) :=
\epsilon(\psi^+, \phi^+) \epsilon(\psi^-, \phi^-)$, and the antilinear map
$J: \spinsp_{\genjpm^-, \genjpm^+} \rightarrow  \spinsp_{\genjpm^-, \genjpm^+}$ by
$J: \psi^+ \otimes \psi^- \mapsto (J\psi^+) \otimes (J\psi^-)$, so that
\begin{equation}
\label{fullJeq}
\epsilon(\Psi, \Phi)=\langle J \Psi, \Phi \rangle.
\end{equation}
As in the case of the $SU(2)$ representations, all
$Spin(4)$ representation operators commute with $J$,
and all generators anticommute with $J$.
Lastly, let $\geniota{}{}$ denote the intertwining map from
$\spinsp_{\rotk}$ to $\spinsp_{\genjpm^-} \otimes \spinsp_{\genjpm^+}$, scaled
such that it is isometric in the Hilbert space inner products.
%
%

\subsection{Canonical phase space, kinematical quantization, and the EPRL vertex}
\label{cansect}

In the general boundary formulation of quantum mechanics \cite{rovelli2004},
one associates to the boundary of any 4-dimensional region a \textit{phase space},
whose quantization yields the \textit{boundary Hilbert space} of the theory for that region.
In the present case, the region is the 4-simplex $\simp$.
The boundary data consists in the algebra elements $B_{ab}$ and
$J_{ab}$ in the frame
of each tetrahedron $a$, and for each pair of tetrahedra $a,b$ one has
a parallel transport map $G_{ab}$ from $b$ to $a$,
related to the $G_a$ introduced in section
\ref{clframe} by $G_{ab} = (G_a)^{-1} G_b$.
These boundary data form a classical phase space isomorphic to
the cotangent bundle over any choice of ten independent parallel transport maps $G_{ab} = (X_{ab}^+,X_{ab}^-)$,
$\Gamma = T^*(Spin(4)^5) = T^*((SU(2) \times SU(2))^5)$, which for simplicity we
choose to be the ten with $a<b$.
For $a<b$, $J_{ab}=(J_{ab}^-,J_{ab}^+)$ and $J_{ba}=(J_{ba}^-,J_{ba}^+)$ respectively
generate right and left translations on $G_{ab}$.

The boundary Hilbert space of states $\Hil_{\partial\simp}^{Spin(4)}$ is the $L^2$ space over the ten
$G_{ab}=(X_{ab}^-, X_{ab}^+) \in Spin(4)$ with $a<b$.
The momenta operators $\hat{J}_{ab}^\pm$ and $\hat{J}_{ba}^\pm$ then act
by $i$ times right and left invariant vector fields, respectively, on the
elements $X_{ab}^\pm$,  and, in terms of these,
$\hat{\rotgen}_{ab}^i := (\hat{J}_{ab}^-)^i + (\hat{J}_{ab}^+)^i$.
One can define an overcomplete basis of $\Hil_{\partial \simp}^{Spin(4)}$,
the \textit{projected spin-network states} (see \cite{livine2002, alexandrov2007}),
each element of which is labeled by four spins $\genjpm_{ab}^\pm, \rotk_{ab}, \rotk_{ba}$ and
two states $\psi_{ab} \in V_{\rotk_{ab}}, \psi_{ba} \in V_{\rotk_{ba}}$ per triangle:
%
%
\begin{equation}
\label{projsn}
\Psi_{\{\genjpm_{ab}^\pm, \rotk_{ab}, \psi_{ab}\}}(G_{ab})
:= \prod_{a<b} \epsilon( \geniota{ab}{ab}\psi_{ab},
\rho(G_{ab}) \geniota{ba}{ab}\psi_{ba}) .
\end{equation}
When acting on such a state, the operators $\hat{L}^i_{ab}, \hat{L}^i_{ba}$
act specifically on the irreducible representation (irrep) vectors
$\psi_{ab}, \psi_{ba}$:
\begin{eqnarray}
\label{leftL_projsn}
\hat{L}^i_{ab} \Psi_{\{j^\pm_{cd},k_{cd},\psi_{cd}\}}
\hspace{-0.2cm} &=& \hspace{-0.2cm}
\epsilon( \geniota{ab}{ab} \hat{L}^i \psi_{ab},
\rho(G_{ab}) \geniota{ba}{ab}\psi_{ba})
\hspace{-0.4cm} \prod_{c<d, (cd) \neq (ab)}  \hspace{-0.4cm}
\epsilon( \geniota{cd}{cd}\psi_{cd},
\rho(G_{cd}) \geniota{dc}{cd}\psi_{dc}) , \\
\label{rightL_projsn}
\hat{L}^i_{ba} \Psi_{\{j^\pm_{cd},k_{cd},\psi_{cd}\}}
\hspace{-0.2cm}&=& \hspace{-0.2cm}
\epsilon( \geniota{ab}{ab} \psi_{ab},
\rho(G_{ab}) \geniota{ba}{ab}  \hat{L}^i \psi_{ba})
\hspace{-0.4cm} \prod_{c<d, (cd) \neq (ab)}  \hspace{-0.4cm}
\epsilon( \geniota{cd}{cd}\psi_{cd},
\rho(G_{cd}) \geniota{dc}{cd}\psi_{dc}) .
\end{eqnarray}
In terms of the projected spin-network overcomplete basis,
the linear simplicity constraint, when quantized as in \cite{elpr2007},
is equivalent to
\begin{equation}
\label{finalsimp}
\rotk_{ab} = \frac{2 \genjpm^-_{ab}}{|1-\gamma|} = \frac{2 \genjpm^+_{ab}}{|1+\gamma|}
= \rotk_{ba}
\end{equation}
for all $a \neq b$.
The projected spin networks satisfying linear simplicity
are  thus  parameterized
by one spin $\rotk_{ab}$ and two states $\psi_{ab}, \psi_{ba} \in V_{\rotk_{ab}}$ per triangle $(ab)$,
the same parameters specifying a generalized
\textit{$SU(2)$ spin-network state} of LQG:
\begin{equation}
\label{spinnet}
\Psi_{\{\rotk_{ab},\psi_{ab}\}}(X_{ab}) := \prod_{a < b} \epsilon( \psi_{ab}, \rho(X_{ab}) \psi_{ba} ) \in \Hil^{LQG}_{\partial\simp}
\equiv L^2(SU(2)^{10}) .
\end{equation}
Because $\genjpm^\pm_{ab} = \half |1 \pm \gamma| \rotk_{ab}$ are always
half-integers, one deduces that only certain values of the spins $k_{ab}$ are allowed;
let $\spinset$ be this set of allowable values, and let $\Hil_{\partial \simp}^{\gamma}$ be the span
of the $SU(2)$ spin-networks (\ref{spinnet}) with $\{\rotk_{ab}\} \subset \spinset$.
One has an embedding
\begin{eqnarray}
\nonumber
\iota: \Hil^{\gamma}_{\partial \simp} & \rightarrow & \Hil^{Spin(4)}_{\partial \simp}\\
\label{EPRLiota}
\Psi_{\{\rotk_{ab}, \psi_{ab}\}} & \mapsto & \Psi_{\{\simpj_{ab}^\pm, \rotk_{ab}, \psi_{ab}\}}
\end{eqnarray}
where here, and throughout the rest of the paper, we set
\begin{equation}
\label{skrelation}
\simpj^\pm := \half|1\pm \gamma|\rotk.
\end{equation}
Due to (\ref{leftL_projsn}) and (\ref{rightL_projsn})
(and because the $SU(2)$ spin-networks satisfy a similar property),
this embedding in fact \textit{intertwines} the spatial rotation generators $\hat{L}_{ab}^i$
in the $Spin(4)$ and $SU(2)$ theories.
Through the embedding $\iota$, the
operators $\hat{L}^i_{ab}$ in the $SU(2)$ theory thus have the same physical meaning
as the corresponding operators in the $Spin(4)$ boundary theory.

Having reviewed the above, the EPRL vertex for a given LQG boundary state $\Psi_{\{k_{ab}, \psi_{ab}\}}^{LQG} \in \Hil_{\partial \simp}^{\gamma} \subset \Hil_{\partial \simp}^{LQG}$
is then
\begin{eqnarray}
\nonumber
A_v(\{k_{ab}, \psi_{ab}\}) &:=&
A_v(\Psi_{\{k_{ab}, \psi_{ab}\}})
= \int_{\rm Spin(4)^5} \prod_a \dif G_a
(\iota \Psi_{\{k_{ab},\psi_{ab}\}})(G_{ab}) \\
\label{eprl}
&=& \int_{\rm Spin(4)^5} \prod_a \dif G_a
\prod_{a<b} \epsilon(\simpiota{ab} \psi_{ab}, \rho(G_{ab})
\simpiota{ab} \psi_{ba}) .
\end{eqnarray}

\section{Proposed proper EPRL vertex}

\subsection{Definition}
Let us consider the structure of the
original EPRL vertex amplitude (\ref{eprl}):
The integration over the group elements $G_a$
can, in a precise sense, be interpreted
as a ``sum over histories'' of parallel transports
from the tetrahedra frames to the 4-simplex frames.
This integration over the $G_a$'s
inside the vertex amplitude can be thought of as a
remnant of the process of integrating out the discrete
connection used to obtain the initial BF spin-foam
model (see \cite{baez1999}).
%
%
Furthermore, in the semiclassical analysis
\cite{bdfgh2009}, one sees that
the $G_a$'s over which one integrates in (\ref{eprl})
play precisely the role of such parallel transports.
Given this interpretation of the $G_a$'s, in order to impose the desired restriction
to the Einstein-Hilbert sector, one must
restrict the discrete history data $G_a$ so that they satisfy
the inequality (\ref{EHineq}):
\begin{equation}
\beta_{ab}(\{G_{a'b'}\}) \tr(\tau_i X^-_{ab} X^+_{ba}) L^i_{ab} > 0 .
\end{equation}
Normally one would do this by inserting
into the path integral
\begin{equation}
\label{clfactor}
\Theta(\beta_{ab}(\{G_{a'b'}\}) \tr(\tau_i X^-_{ab} X^+_{ba}) L^i_{ab})
\end{equation}
where $\Theta$ is the Heaviside function, defined to be zero when its argument is zero.
However, in the integral (\ref{eprl}),
it is not the classical
quantity $L^i_{ab}$ that appears, but rather
\textit{states} $\psi_{ab}$ in irreducible
representations of the corresponding operators
$\hat{L}^i_{ab}$.\footnote{If
one uses coherent boundary data as will be done in
the next section, then one does have a classical
label $L^i_{ab}$ present, but one would still not
be able to simply insert the factor (\ref{clfactor}), as,
due to the overcompleteness of the set of coherent
states, this would lead to a vertex amplitude that is not linear
in the boundary state, something necessary to ensure the final transition amplitude defined by the spin-foam sum is
linear in the boundary state.
}
As noted in equations (\ref{leftL_projsn}) and
(\ref{rightL_projsn}), $\hat{L}^i_{ab}$ acts on
$\psi_{ab}$ via the $SU(2)$ generators $\hat{L}^i$.
Therefore, we partially `quantize' the expression
 (\ref{clfactor}) by replacing $L^i_{ab}$ with
the generators $\hat{L}^i$, yielding the following
$G_a$-dependent
operator acting
in the spin $k_{ab}$ representation of $SU(2)$:
\begin{equation}
\label{Pabdef}
P_{ba}(\{G_{a'b'}\}):= P_{(0,\infty)}
\left(\beta_{ab}(\{G_{a'b'}\})
\tr(\tau_i X^-_{ba}X^+_{ab})\hat{L}^i\right),
\end{equation}
where $P_{\scr{S}}(\hat{O})$ denotes the spectral projector onto the portion $\scr{S} \subset \R$
of the spectrum of the operator $\hat{O}$.
Inserting (\ref{Pabdef}) into the face factors of (\ref{eprl}) we obtain what we call the
\textit{proper EPRL vertex amplitude}:
\begin{equation}
\label{propeprl}
A^{\prop}_v(\{k_{ab}, \psi_{ab}\}) :=
\int_{\rm Spin(4)^5} \prod_a \dif G_a
\prod_{a<b} \epsilon(\simpiota{ab} \psi_{ab}, \rho(G_{ab})
\simpiota{ab} P_{ba}(\{G_{a'b'}\}) \psi_{ba}) .
\end{equation}
Let us stop for a moment and remark on the properties of this vertex
amplitude. First, as the EPRL vertex, it depends on an $SU(2)$ spin network boundary
state and hence may be used to construct a spin-foam model
\textit{for loop quantum gravity}. It is linear in the SU(2) boundary state, as required for the final spin-foam
amplitude to be linear in the initial state and antilinear in the final state.
Furthermore, as we will show
in the next subsection, it is invariant under
SU(2) gauge transformations.
Finally, and most importantly,
as we  will show in the next section, its asymptotics only include the single term
$e^{i S_{\Regge}}$, as desired.

Throughout the rest of this paper,
the notation $P_{ba}(\{G_{a'b'}\})$
introduced in (\ref{Pabdef}) will
also refer to the projector
acting in the spin $(\simpj^-_{ab}, \simpj^+_{ab})$ representation of $Spin(4)$, defined by the same
expression (\ref{Pabdef}). In each statement using the notation $P_{ba}(\{G_{a'b'}\})$,
either the context will determine which projector is intended, or the statement will hold for both projectors.

Finally, let us briefly note two ways to rewrite the proper vertex:
(1.)
It may at first appear arbitrary that the projector was inserted on the right side of each
face factor in equation (\ref{propeprl}).  However, in fact, one can put the projector
(appropriately transformed) anywhere in each face-factor, and the vertex amplitude
doesn't change.  See appendix \ref{leftapp}.
(2.)
We note that, using equation (\ref{fullJeq}),
one has the following equivalent expression for the
proper vertex:
\begin{equation}
\label{propeprl_herm}
A^{\prop}_v(\{k_{ab}, \psi_{ab}\}) :=
\int_{\rm Spin(4)^5} \prod_a \dif G_a
\prod_{a<b} \langle J \simpiota{ab} \psi_{ab}, \rho(G_{ab})
\simpiota{ab} P_{ba}(\{G_{a'b'}\}) \psi_{ba} \rangle .
\end{equation}

\subsection{Proof of invariance under $SU(2)$ gauge transformations}

\begin{theorem}
The proper EPRL vertex is invariant under arbitrary $SU(2)$ gauge transformations at the tetrahedra.
\end{theorem}
{\startproof
Let $\{\rotk_{ab}, \psi_{ab}\}$ be the data for a given spin network on the boundary, and let five $SU(2)$ elements
$h_a$, one at each tetrahedron, be given.
We wish to show \\
$A_v^\prop(\Psi_{\{\rotk_{ab},  \rho(h_a)\psi_{ab}\}}) = A_v^\prop(\Psi_{\{\rotk_{ab}, \psi_{ab}\}})$.

First, define $\tilde{G}_{ab}:= (h_a, h_a)^{-1} \circ G_{ab} \circ (h_{b}, h_{b})$.   Then
\begin{eqnarray}
\nonumber
\left(\tilde{G}_{ab} \cdot \scr{N}\right)^i &=&
\tr(\tau^i \tilde{X}_{ab}^- \tilde{X}_{ba}^+)
= \tr(\tau^i h_{a}^{-1} X_{ab}^- X_{ba}^+ h_a) \\
\nonumber
&=& \tr((h_a \tau^i h_{a}^{-1}) X_{ab}^- X_{ba}^+)
=  h_a \triangleright \tr(\tau^i  X_{ab}^- X_{ba}^+) \\
\label{Gncov}
&=& h_a \triangleright \left( G_{ab} \cdot \scr{N}\right)^i .
\end{eqnarray}
From this and the $SO(3)$ invariance of $\epsilon_{ijk}$, it follows that
\begin{equation}
\label{betainv}
\beta_{ab}(\{\tilde{G}_{a'b'} \}) = \beta_{ab}(\{G_{a'b'} \}) .
\end{equation}
We thus have
\begin{eqnarray}
\nonumber
\rho(h_b)^{-1} P_{ba}(\{G_{a'b'}\}) \rho(h_b)
&=& \rho(h_b)^{-1} P_{(0,\infty)}\left( \beta_{ab}(\{G_{a'b'}\}) (G_{ba} \cdot \scr{N})_i \hat{L}^i\right) \rho(h_b) \\
\nonumber
&=& P_{(0,\infty)}\left( \beta_{ab}(\{G_{a'b'}\}) [(h_b)^{-1} \triangleright (G_{ba} \cdot \scr{N})_i] \hat{L}^i\right) \\
\nonumber
&=& P_{(0,\infty)}\left( \beta_{ab}(\{\tilde{G}_{a'b'}\}) (\tilde{G}_{ba} \cdot \scr{N})_i \hat{L}^i\right) \\
\label{Prot}
&=& P_{ba}(\{\tilde{G}_{a'b'}\})
\end{eqnarray}
where lemma \ref{projgroup_comm}  has been used in the second line, and (\ref{Gncov}) and (\ref{betainv}) have been used in the third. Using (\ref{Prot}), we finally have
\begin{eqnarray*}
\dummy \hspace{2.2cm}&& \hspace{-3cm} A^\prop_v(\{k_{ab}, \rho(h_a) \psi_{ab}\})
:= \int \left(\prod_{a<b} \dif G_{ab}\right)
\prod_{a<b} \epsilon\left(\simpiota{ab} \rho(h_a) \psi_{ab},
\rho(G_{ab}) \simpiota{ab}  P_{ba}(\{G_{a'b'}\}) \rho(h_b) \psi_{ba}\right) \\
&=& \!\!\! \int \left(\prod_{a<b} \dif G_{ab}\right)
\prod_{a<b} \epsilon\left(\simpiota{ab} \rho(h_a) \psi_{ab},
\rho(G_{ab}) \simpiota{ab}  \rho(h_b)  P_{ba}(\{\tilde{G}_{a'b'}\})\psi_{ba}\right) \\
&=& \!\!\! \int \left(\prod_{a<b} \dif G_{ab}\right)
\prod_{a<b} \epsilon\left(\simpiota{ab} \psi_{ab},
\rho(h_a,h_a)^{-1}\rho(G_{ab}) \rho(h_b, h_b)\simpiota{ab}   P_{ba}(\{\tilde{G}_{a'b'}\})\psi_{ba}\right) \\
&=& \!\!\! \int \left(\prod_{a<b} \dif G_{ab}\right)
\prod_{a<b} \epsilon\left(\simpiota{ab} \psi_{ab},
\rho(\tilde{G}_{ab})\simpiota{ab}   P_{ba}(\{\tilde{G}_{a'b'}\})\psi_{ba}\right) \\
&=& \!\!\! \int \left(\prod_{a<b} \dif\tilde{G}_{ab}\right)
\prod_{a<b} \epsilon\left(\simpiota{ab} \psi_{ab},
\rho(\tilde{G}_{ab})\simpiota{ab}   P_{ba}(\{\tilde{G}_{a'b'}\})\psi_{ba}\right) \\
&=& \!\!\! A^\prop_v\left(\{k_{ab}, \psi_{ab}\}\right)
\end{eqnarray*}
where we have used in the third line the intertwining property of $\simpiota{ab}$
and in the second to last line the right and left invariance of the Haar measure.
\finishproof}

\subsection{$Spin(4)$ invariance}
%
%

As mentioned in section \ref{clsect}, in defining the classical discrete variables $\{G_a, B_{ab}\}$, one thinks of each tetrahedron as having its own `frame'.
Concretely, this is manifested in the fact that there exists a local $Spin(4)$ gauge transformation acting at each tetrahedron.
Given a choice of $Spin(4)$ group element $H_a$ at each tetrahedron $a$, one has the following gauge transformation:
\begin{equation}
(\{H_{a'}\}) \cdot G_a = G_a H_a,
\qquad (\{H_{a'}\}) \cdot B_{ab} = H_a \triangleright B_{ab} .
\end{equation}
The definition of the proper vertex (\ref{propeprl}) makes key use of a fixed internal direction $\mathcal{N}^I = (1,0,0,0)$.
This vector is used to impose the simplicity constraints (\ref{gfconstr}) at each tetrahedron, and superficially breaks the above $Spin(4)$ gauge symmetry.
Furthermore, in order to embed LQG states
into BF states solving simplicity, the proper vertex uses the map $\simpiota{}$, which is defined using a specific embedding
$h: g \mapsto (g,g)$ of $SU(2)$ into $Spin(4)$ via the symmetry condition
\begin{equation}
\label{iotasymm}
\simpiota{} \circ \rho (g) = \rho(h(g,g)) \circ \simpiota{}.
\end{equation}
This use of $h$ also seems to break the above $Spin(4)$ symmetry.
The fixed vector $\mathcal{N}^I$ and embedding $h$ are related by the fact that
the $SO(4)$ action of every element in the image of $h$ preserves $\mathcal{N}^I$.
(The original EPRL vertex amplitude uses these two exact same extra structures \cite{epr2007a, elpr2007}.)

$Spin(4)$ acts on the unit vector $\mathcal{N}^I$ by its $SO(4)$ action, while it acts on the map
$\simpiota{}$ via
\begin{equation}
(\Lambda \cdot \iota)_k^{s^-, s^+}:= \rho(\Lambda) \circ \geniota{}{}
\end{equation}
for $\Lambda \in Spin(4)$.
The transformed map $(\Lambda \cdot \iota)_k^{j^-, j^+}: V_k \rightarrow V_{j^-, j^+}$ still
satisfies a symmetry condition similar to (\ref{iotasymm}), but with a different embedding $(\Lambda \cdot h): SU(2) \rightarrow Spin(4)$:
\begin{equation}
(\Lambda \cdot \iota)_k^{s^-, s^+} \circ \rho(g) = \rho((\Lambda \cdot h)(g)) \circ (\Lambda \cdot \iota)_k^{s^-, s^+}
\end{equation}
where $(\Lambda \cdot h)(g):= \Lambda h(g) \Lambda^{-1}$.

In this section we consider what happens when, in the definition of the proper vertex,
the unit vector $\mathcal{N}^I$ and the map $\simpiota{}$ are replaced, at each tetrahedron $a$,
by their transformation under an arbitrary $Spin(4)$ element $\Lambda_a$.
The resulting, a priori possibly modified proper vertex amplitude we denote by ${}^{\{\Lambda_a\}}A^{\prop}_v$.
An arbitrary $Spin(4)$ gauge transformation $\{H_a\}$ then acts on  ${}^{\{\Lambda_a\}}A^{\prop}_v$ via
\begin{equation}
{}^{\{\Lambda_a\}} A^{\prop}_v \mapsto {}^{\{H_{a} \Lambda_a \}}A^{\prop}_v .
\end{equation}
We shall prove that the generalized proper vertex  ${}^{\{\Lambda_a\}}A^{\prop}_v$ is in fact independent of
$\{ \Lambda_a \}$, and so is trivially invariant under the above action and in this sense is $Spin(4)$ invariant at each
tetrahedron.  This result is similar to that in \cite{rs2010a}.

We begin by noting how to write $A^{\prop}_v$ in a way that makes its dependence
on $\mathcal{N}^I$ explicit, which then allows us to write down explicitly the generalized proper vertex
${}^{\{\Lambda_a\}}A^{\prop}_v$, after which we prove its independence of $\{\Lambda_a\}$.
From the first line of equation (\ref{leftexp}),
\begin{equation}
A^{\prop}_v(\{k_{ab}, \psi_{ab}\})
= \int_{\rm Spin(4)^5} \prod_a \dif G_a
\prod_{a<b} \epsilon(\simpiota{ab} \psi_{ab}, \rho(G_{ab}) P_{ba}(\{G_{a'b'}\})
\simpiota{ba} \psi_{ba}) .
\end{equation}
The above projector $P_{ba}(\{G_{a'b'}\})$ on $V_{s^-_{ab}, s^+_{ab}}$ can be written
\begin{equation}
P_{ba}(\{G_{a'b'}\}) := P_{(0,\infty)}
(\beta_{ba}(\{G_{a'b'}\})\epsilon_{IJKL} \mathcal{N}^I (G_{ba}\cdot \mathcal{N})^J \hat{J}^{KL})
\end{equation}
with
\begin{eqnarray*}
\dummy \hspace{-0.3cm}\beta_{ba}(\{G_{a'b'}\}) &:=&
 -\sgn\left[\epsilon_{IJKL} \scr{N}^I (G_{ac}\cdot\scr{N})^J  (G_{ad}\cdot\scr{N})^K  (G_{ae}\cdot\scr{N})^L \right. \cdot \\
&& \hspace{0.8cm} \left. \cdot \epsilon_{MNPQ} \scr{N}^M (G_{bc}\cdot\scr{N})^N  (G_{bd}\cdot\scr{N})^P  (G_{be}\cdot\scr{N})^Q
\right]
\end{eqnarray*}
with $\{c,d,e\} = \{0, \dots, 4 \} \setminus \{a,b\}$ in any order.
This immediately yields the following expression for the generalized proper vertex:
\begin{equation}
\label{genprop}
\genprop{a'} (\{k_{ab}, \psi_{ab}\})
= \int_{\rm Spin(4)^5} \prod_a \dif G_a
\prod_{a<b} \epsilon(\rho(\Lambda_a) \simpiota{ab} \psi_{ab}, \rho(G_{ab}) \genproj{a'}{ba}(\{G_{a'b'}\})
\rho(\Lambda_b)\simpiota{ba} \psi_{ba}) .
\end{equation}
where
\begin{equation}
\genproj{a'}{ba}(\{G_{a'b'}\}) := P_{(0,\infty)}
( \genbeta{a'}{ba}(\{G_{a'b'}\})\epsilon_{IJKL} (\Lambda_b \cdot \mathcal{N})^I
 (G_{ba} \Lambda_a \cdot \mathcal{N})^J \hat{J}^{KL})
\end{equation}
with
\begin{eqnarray*}
\dummy \hspace{-0.3cm} \genbeta{a'}{ba}(\{G_{a'b'}\}) &:=&
 -\sgn\left[\epsilon_{IJKL} (\Lambda_a \cdot \scr{N})^I (G_{ac} \Lambda_c \cdot\scr{N})^J
(G_{ad} \Lambda_d \cdot\scr{N})^K  (G_{ae} \Lambda_e \cdot\scr{N})^L \right. \cdot \\
&& \hspace{0.8cm} \left. \cdot \epsilon_{MNPQ} (\Lambda_b \cdot \scr{N})^M (G_{bc}\Lambda_c \cdot\scr{N})^N
(G_{bd}\Lambda_d \cdot\scr{N})^P  (G_{be}\Lambda_e \cdot\scr{N})^Q
\right]
\end{eqnarray*}
%
%
\begin{theorem}
$\genprop{a'} (\{k_{ab}, \psi_{ab}\}) = A^{\prop}_v(\{k_{ab}, \psi_{ab}\})$
for all $\{\Lambda_{a'}\} \subset Spin(4)$.
\end{theorem}
{\startproof

Let $\tilde{G}_a := \Lambda_a G_a$.
We begin by proving 
(i.) $\genbeta{a'}{ba}(\{G_{ab}\}) = \beta_{ba}(\{\tilde{G}_{ab}\})$,
and
(ii.) $\genproj{a'}{ba}(\{G_{a'b'}\})
= \rho(\Lambda_b) \circ P_{ba}(\{\tilde{G}_{a'b'}\}) \circ \rho(\Lambda_b)^{-1}$.
Using these facts in (\ref{genprop}), together with the $Spin(4)$ invariance of the $\epsilon$-inner product on $V_{j^-, j^+}$ and
right invariance of the Haar measure, then yields the result.

(i.)

\newcommand{\iispace}{\hspace{0.9cm}}
\begin{eqnarray*}
\dummy \hspace{-0.2cm}\genbeta {a'}{ba}(\{G_{a'b'}\})
&:=&
 -\sgn\left[\epsilon_{IJKL} (\Lambda_a \cdot \scr{N}_a)^I (G_{ac}\Lambda_c\cdot\scr{N}_c)^J  (G_{ad}\Lambda_d\cdot\scr{N}_d)^K  (G_{ae}\Lambda_e\cdot\scr{N}_e)^L \cdot \right. \\
&& \iispace
\left. \cdot \epsilon_{MNPQ} (\Lambda_b \cdot \scr{N}_b)^M (G_{bc}\Lambda_c\cdot\scr{N}_c)^N
(G_{bd}\Lambda_d\cdot\scr{N}_d)^P  (G_{be}\Lambda_e\cdot\scr{N}_e)^Q
\right] \\
&=&
 -\sgn\left[\epsilon_{IJKL} (\Lambda_a \cdot \scr{N}_a)^I (\Lambda_a \tilde{G}_{ac}\cdot\scr{N}_c)^J  (\Lambda_a \tilde{G}_{ad}\cdot\scr{N}_d)^K
(\Lambda_a \tilde{G}_{ae}\cdot\scr{N}_e)^L \cdot \right. \\
&& \iispace
\left. \cdot \epsilon_{MNPQ} (\Lambda_b \cdot \scr{N}_b)^M (\Lambda_b \tilde{G}_{bc}\cdot\scr{N}_c)^N  (\Lambda_b \tilde{G}_{bd}\cdot\scr{N}_d)^P
(\Lambda_b \tilde{G}_{be}\cdot\scr{N}_e)^Q
\right] \\
&=&
 -\sgn\left[\epsilon_{IJKL} \scr{N}_a^I (\tilde{G}_{ac}\cdot\scr{N}_c)^J  (\tilde{G}_{ad}\cdot\scr{N}_d)^K
(\tilde{G}_{ae}\cdot\scr{N}_e)^L \right. \cdot \\
&& \iispace
\left. \cdot \epsilon_{MNPQ} \scr{N}_b^M (\tilde{G}_{bc}\cdot\scr{N}_c)^N  (\tilde{G}_{bd}\cdot\scr{N}_d)^P
(\tilde{G}_{be}\cdot\scr{N}_e)^Q
\right] \\
&=& \beta_{ba}(\{\tilde{G}_{a'b'}\})
\end{eqnarray*}
where the $SO(4)$ invariance of $\epsilon_{IJKL}$ was used.
%
%

(ii.)

\begin{eqnarray*}
\genproj{a'}{ba}(\{G_{a'b'}\}) &:=& P_{(0,\infty)}(\beta_{ba}(\{\tilde{G}_{a'b'}\})
\epsilon_{IJKL} (\Lambda_b \cdot \mathcal{N})^I (G_{ba} \Lambda_a \cdot \mathcal{N})^J \hat{J}^{KL}) \\
&=& P_{(0,\infty)} (\beta_{ba}(\{\tilde{G}_{a'b'}\})
\epsilon_{IJKL} (\Lambda_b \cdot \mathcal{N})^I (\Lambda_b \tilde{G}_{ba} \cdot \mathcal{N})^J \hat{J}^{KL}) \\
&=& P_{(0,\infty)} (\beta_{ba}(\{\tilde{G}_{a'b'}\})
\epsilon_{IJKL} \mathcal{N}^I (\tilde{G}_{ba} \cdot \mathcal{N})^J (\Lambda_b^{-1})^K{}_M  (\Lambda_b^{-1})^L{}_N  \hat{J}^{MN}) \\
&=& P_{(0,\infty)} (\rho(\Lambda_b) \beta_{ba}(\{\tilde{G}_{a'b'}\})
\epsilon_{IJKL} \mathcal{N}^I (\tilde{G}_{ba} \cdot \mathcal{N})^J \hat{J}^{MN} \rho(\Lambda_b)^{-1}) \\
%
%
&=&  \rho(\Lambda_b) \circ P_{(0,\infty)} (\beta_{ba}(\{\tilde{G}_{a'b'}\})
\epsilon_{IJKL} \mathcal{N}^I (\tilde{G}_{ba} \cdot \mathcal{N})^J \hat{J}^{MN}) \circ  \rho(\Lambda_b)^{-1} \\
&=&  \rho(\Lambda_b) \circ P_{ba}(\{\tilde{G}_{a'b'}\}) \circ  \rho(\Lambda_b)^{-1}
\end{eqnarray*}
where result (i.) was used in the first line, the $SO(4)$ invariance of $\epsilon_{IJKL}$ in the third line, and the 
$Spin(4)$ covariance of the generators $\hat{J}^{KL}$ in the fourth line.
\finishproof }

\subsection{Lorentzian generalization}
\label{lorsect}

\newcommand{\Lor}{\text{Lor}}

We close this section by noting that there is an obvious generalization of the expression (\ref{propeprl}) of the proper vertex to the Lorentzian signature.
In the Lorentzian EPRL model \cite{elpr2007, pereira2007}, one uses the unitary representations of $SL(2,\C)$, which are labeled by a real number $\rho$ together with 
an integer $n$.  Denote the carrying space for such representations by $V_{\rho, n}^\Lor$, and let $\rho(G)$ denote the representation thereon of $G \in SL(2,\C)$.
$V_{\rho, n}^\Lor$ decomposes into an infinite direct sum of irreducible representations of $SU(2)$:
\begin{equation}
\label{lordecomp}
V_{\rho, n}^\Lor = \oplus_{k=n/2}^\infty V_k
\end{equation}
where in the sum $k$ is incremented in steps of $1$.  The analogue of the embedding $\simpiota{}: V_k \rightarrow V_{s^-, s^+}$ in the Lorentzian case
is the embedding $\mathcal{I}_k : V_k \rightarrow V_{2 \gamma k, 2k}^\Lor$ mapping $V_k$ into the lowest $k$ component 
of $V_{2\gamma k, 2k}^\Lor$ in the sum (\ref{lordecomp}).
The elements in the image of this embedding satisfy a quantization of the simplicity constraints just as those of $\simpiota{}{}$ 
do in the Euclidean case \cite{elpr2007}.
Furthermore, just as one has the invariant bilinear form $\epsilon$ on $V_{s^-, s^+}$, related to the Hermitian inner product via the antilinear map $J$, so too
one has an invariant bilinear form $\beta$ on $V^\Lor_{\rho, n}$, related to the Hermitian inner product on $V^\Lor_{\rho, n}$ via an antilinear map $\scr{J}$
in the same way \cite{bdfhp2009}. 
%
%
For simple representations, $(j^+, j^-) = (s^-, s^+)$, $(\rho, n) = (2\gamma k, 2k)$,
$\epsilon$ and $\beta$ furthermore have the same (anti-)symmetry properties: $\epsilon(\psi,\phi) = (-1)^{2k} \epsilon(\phi, \psi)$,
$\beta(\psi, \phi) = (-1)^{2k} \beta(\phi, \psi)$.  In terms of these structures, the expression for the Lorentzian EPRL vertex amplitude is exactly analogous to the Euclidean expression (\ref{eprl})
\cite{elpr2007, bdfhp2009}:
\begin{equation}
A^\Lor_v(\{k_{ab}, \psi_{ab}\}) = \int_{\rm SL(2,\C)^4} \prod_{a\neq 4} \dif G_a
\prod_{a<b} \beta(\scr{I}_{k_{ab}} \psi_{ab}, \rho(G_{ab})
\scr{I}_{k_{ab}} \psi_{ba}) ,
\end{equation}
the only notable difference being that one of the group integrations is dropped in order to ensure finiteness of the amplitude \cite{bb2001, ep2008}.
One can then modify this vertex amplitude in exactly the same way as was done in the Euclidean case, to yield a Lorentzian version of the proper EPRL vertex:
\begin{equation}
\label{lorprop}
A^{\prop, \Lor}_v(\{k_{ab}, \psi_{ab}\}) :=
\int_{\rm SL(2,\C)^4} \prod_{a\neq 4} \dif G_a
\prod_{a<b} \beta(\scr{I}_{k_{ab}} \psi_{ab}, \rho(G_{ab})
\scr{I}_{k_{ab}} P_{ba}(\{G_{a'b'}\}) \psi_{ba}) .
\end{equation}
where
\begin{equation}
P_{ba}(\{G_{a'b'}\}):= P_{(0,\infty)}
\left(\beta_{ab}(\{G_{a'b'}\})
(G_{ba} \cdot \scr{N})_i \hat{L}^i\right),
\end{equation}
with
\begin{equation}
\dummy \hspace{-0.2cm}\beta_{ab}(\{G_{a'b'}\}):= -\sgn\left[\epsilon_{ijk} (G_{ac}\cdot\scr{N})^i  (G_{ad}\cdot\scr{N})^j  (G_{ae}\cdot\scr{N})^k
\epsilon_{lmn} (G_{bc}\cdot\scr{N})^l  (G_{bd}\cdot\scr{N})^m  (G_{be}\cdot\scr{N})^n\right]
\end{equation}
with $\{c,d,e\} = \{0, \dots, 4 \} \setminus \{a,b\}$ in any order,
and where $G \cdot \scr{N}$ denotes the $SO(1,3)$ action of $G$ on $\scr{N}^I = (1,0,0,0)$.  Though this generalization of the proper vertex to the Lorentzian signature is natural, and it 
is difficult to imagine how otherwise to generalize to this case, nevertheless one should justify this generalization more systematically, by quantizing an appropriate classical condition isolating the Lorentzian Einstein-Hilbert sector.
One should also check whether the above generalization has the required semiclassical limit, as we will prove below is the case for the Euclidean proper vertex.
Henceforth in this paper, unless otherwise indicated, ``proper vertex'' shall again always refer to ``Euclidean proper vertex.''

\section{Asymptotics}

In the following we state and prove the asymptotics of the proper vertex, using key results from
\cite{bdfgh2009}.  

\subsection{Statement of the formula}

It will be useful for later purposes to define the following before
defining coherent states.
\begin{definition}
Given any unit $n^i \in \R^3$, let $\njm{n}{\rotk}{m}$ denote the eigenstate of $n \cdot \hat{L}$ in $V_k$
with eigenvalue $m$, and $\njm{n}{\genjpm^-, \genjpm^+, \rotk}{m}$ the eigenstate of $\hat{L}^2$
and $n \cdot \hat{L}$ in $V_{\genjpm^-. \genjpm^+}$ with eigenvalues $k(k+1)$ and $m$, with phase fixed arbitrarily
for each set of labels.
\end{definition}

\begin{definition}
Given a unit 3-vector $n$, a spin $\genj$, and a phase $\theta$, we define
the corresponding \textit{coherent state} as
\begin{equation}
|n,\theta \rangle_\genj := e^{i\theta}\njm{n}{\genj}{\genj} .
\end{equation}
The $\theta$ argument represents a phase freedom, and will usually
be suppressed.  Additionally,  when the spin is clear from the context, 
it will be omitted. Such coherent states were first used in quantum gravity by Livine and
Speziale \cite{ls2007}.
\end{definition}

We call an assignment of one spin $k_{ab} \in \spinset$ and two unit
3-vectors $n_{ab}^i, n_{ba}^i$ to each triangle $(ab)$ in $\simp$
a set of \textit{quantum boundary data}.  Given such data, the corresponding
boundary \textit{state} in the $SU(2)$ boundary Hilbert space of $\simp$ is
\begin{equation}
\label{genstate}
\Psi_{\{\rotk_{ab},n_{ab}\}, \theta} := \Psi_{\{\rotk_{ab}, \psi_{ab}\}}
\qquad \text{with} \qquad
|\psi_{ab}\rangle := |n_{ab}, \theta_{ab} \rangle_{\rotk_{ab}}
\end{equation}
where the $\theta_{ab}$ are any phases summing to $\theta$ modulo $2\pi$. The phase $\theta$ will usually
be suppressed.
%
%
The state $\Psi_{\{\rotk_{ab},n_{ab}\}}$ so defined is a coherent boundary state
corresponding to the classical reduced boundary data $A_{ab} = A(k_{ab}) := \kappa \gamma k_{ab}$
and $n_{ab}$.

When $\{A(\rotk_{ab}), \sut{n}_{ab}\}$ is nondegenerate and satisfies closure,
we also say that $\{\rotk_{ab}, \sut{n}_{ab}\}$ is nondegenerate and satisfies closure.
In this case, for each tetrahedron $a$, there exists a geometrical tetrahedron in $\R^3$,
unique up to translations, such that $\{A(\rotk_{ab})\}_{b\neq a}$ and $\{n^i_{ab}\}_{b\neq a}$
are the areas and outward unit normals, respectively, of the four triangular faces, which we
denote by $\{\Delta_{ab}^t\}_{b \neq a}$.
If these five geometrical tetrahedra can be glued together consistently to form a 4-simplex,
we say that the boundary data $\{k_{ab}, n_{ab}\}$ is \textit{Regge-like}.
For such data, there exists a set
of $SU(2)$ elements $\{g_{ab} = g_{ba}^{-1}\}$, unique up to a $\mathbb{Z}_2$ lift ambiguity \cite{bdfgh2009},
such that the adjoint action of each $g_{ab}$ on $\R^3$ maps (1.) $\Delta_{ab}^t$ into $\Delta_{ba}^t$, and (2.) $n_{ba}$ into  $-n_{ab}$.
These group elements can be used to completely remove the phase ambiguity in the boundary state
(\ref{genstate}),
by requiring the phase of
the coherent states to be chosen such that
$g_{ab} |n_{ba} \rangle_{\rotk_{ab}} = J|n_{ab}\rangle_{\rotk_{ab}}$,
where $J$ is as defined in section \ref{notationsec}.
The resulting boundary state $\Psi_{\{k_{ab}, n_{ab}\}}$
is called the \textit{Regge state} determined by $\{k_{ab}, n_{ab}\}$, and is denoted
by $\Psi^{\Regge}_{\{k_{ab},n_{ab}\}}$.

The following theorem, as theorem 1 in \cite{bdfgh2009}, uses the fact that,
because the boundary data $\{k_{ab}, n_{ab}\}$ determine the geometry of all boundary
tetrahedra, it also determines the geometry of the 4-simplex itself \cite{bdfgh2009, connelly1993},
and hence, in particular, the dihedral angles $\Theta_{ab} \in [0, \pi]$ via the equation
$N_a \cdot N_b = \cos \Theta_{ab}$ where $N_a$ and $N_b$ are the
outward pointing normals to the $a$th and $b$th tetrahedra,
respectively.

\begin{theorem}[Proper EPRL asymptotics]
\label{prop_asym_thm}
If $\{\rotk_{ab}, n_{ab}\}$ is boundary data representing a nondegenerate Regge geometry, then, in the limit of large $\lambda$,
\begin{equation}
\label{prop_asym}
A^\prop_v(\Psi^{\rm Regge}_{\lambda \rotk_{ab},n_{ab}}) \sim \lambda^{-12} N
\exp\left(i \sum_{a<b} A(\lambda \rotk_{ab}) \Theta_{ab}\right)
\end{equation}
where $N$ is independent of $\lambda$ and the error term is bounded by a constant times $\lambda^{-13}$.
If $\{\rotk_{ab}, n_{ab}\}$ does not represent a nondegenerate Regge geometry,
then $A^\prop_v(\Psi_{\lambda \rotk_{ab},n_{ab},\theta})$ decays exponentially with large $\lambda$ for any
choice of phase $\theta$.
\end{theorem}

To prove this theorem, in manner similar to \cite{bdfgh2009},
%
%
we cast the proper vertex in appropriate integral
form $A^\prop_v = \int \dif \mu(x)e^{S_{\gamma < 1}(x)}$ and $A^\prop_v = \int \dif \mu(x)e^{S_{\gamma > 1}(x)}$, separately for  the cases $\gamma < 1$ and $\gamma > 1$,
where $S_{\gamma < 1}$ and $S_{\gamma > 1}$ are ``actions''.
We then determine the critical points for each action.
In proving this theorem, we are interested in critical points whose contributions are not exponentially suppressed. For this reason,
we define the term ``critical point'' to mean points where the action is stationary and its real part is \textit{nonnegative}.
If a point in the domain of integration is such that the real part of the action is an absolute maximum
and is nonnegative, we shall say it is a
\textit{maximal point}.
%
%
%
%

\subsection{Integral expressions and critical points}

In the following, whenever we say the words ``critical points'' with no other
qualification, we refer to critical points of the proper EPRL vertex (\ref{propeprl}).

\subsubsection{The case $\gamma < 1$}

The relevant integral form of the proper vertex in this case is
\begin{equation}
\label{gl1vert}
A^\prop_v(\Psi_{\{\rotk_{ab}, n_{ab}\},\theta})
= \int \prod_a \dif G_a  \exp(S_{\gamma < 1})
\end{equation}
where
\begin{equation}
\label{gl1act}
\exp(S_{\gamma < 1})
= \prod_{a<b}
\langle J \simpiota{ab} n_{ab}, \quad
\rho(G_{ab}) \simpiota{ab}
P_{ba}(\{G_{a'b'}\}) n_{ba} \rangle .
\end{equation}
The action $S_{\gamma < 1}$ is, as in \cite{bdfgh2009}, generally complex.
The two conditions that determine critical points are maximality and stationarity.
In both proving the equations for maximality and checking stationarity,
it will be simplest to reuse the
results in\cite{bdfgh2009}.  This will highlight the simplicity of the additional
steps necessary for the present modification.
Recall from \cite{bdfgh2009} that the action for $\gamma < 1$
for the original EPRL model is
\begin{equation}
\label{gl1EPRLact}
\exp(S^\EPRL_{\gamma < 1})
= \prod_{a<b}
\langle J \simpiota{ab} n_{ab}, \quad
\rho(G_{ab})^{-1} \simpiota{ab}
n_{ba} \rangle .
\end{equation}
For the purpose of the following lemmas and the rest of this section, it is convenient to define
a set of group elements together with boundary data
$\{G_a, k_{ab}, n_{ab}\}$ to satisfy \textit{proper orientation}
if, for all $a < b$,
\mbox{$\beta_{ab}(\{G_{a'b'}\}) \tr (\tau_i X^-_{ba} X^+_{ab})n^i_{ba} > 0$}.
%

\begin{lemma}
\label{gl1max}
Given boundary data $\{k_{ab}, n_{ab}\}$, $\{G_a\}$ is a maximal point
of $S_{\gamma < 1}$ iff orientation and
proper orientation are satisfied.
\end{lemma}
{\startproof
From (\ref{gl1act}),
%
%
\begin{subequations}
\begin{eqnarray}
\nonumber
\exp (\Re S_{\gamma < 1}) &=& |\exp (S_{\gamma < 1})| 
= \prod_{a<b} \left| \left\langle J \iota_{k_{ab}} n_{ab}, \rho(G_{ab}) \iota_{k_{ab}} 
P_{ba}(\{G_{a'b'}\}) n_{ba }\right\rangle \right| \\
\nonumber
&\le& \prod_{a<b} || J \iota_{k_{ab}} | n_{ab} \rangle|| \, 
||\rho(G_{ab}) \iota_{k_{ab}} 
P_{ba}(\{G_{a'b'}\}) | n_{ba } \rangle || \\
\label{gl1ineq}
&=& \prod_{a<b} || P_{ba}(\{G_{a'b'}\}) |n_{ba}\rangle || \le 1
\end{eqnarray}
where the Cauchy-Schwarz inequality has been used in the second line, 
the fact that $J$, $\iota_{k_{ab}}$, and $\rho(G_{ab})$ are norm preserving
and that $|| \, |n_{ab}\rangle||=1$  have been used in the last equality, and  
$|| \, |n_{ba}\rangle||=1$ has been used in the last inequality.

We now proceed to prove that $\exp(\Re S_{\gamma < 1})=1$ iff orientation and proper orientation are satisfied.

\noindent($\Leftarrow$) Suppose orientation and proper orientation are satisfied.  
From equation (52) in \cite{bdfgh2009}, it follows that, for each $a\neq b$, there exists $\lambda_{ba}$ such that
$X^-_{ba}X^+_{ab} = \exp (\lambda_{ba} n_{ba} \cdot \tau)$, so that
$\tr (\tau^i X^-_{ba}X^+_{ab}) \widehat{=}: n[X^-_{ba} X^+_{ab}]^i =  \pm n_{ba}^i$. 
Proper orientation then implies that the sign in this equation is $\beta_{ab}(\{G_{a'b'}\})$ for all $a<b$. 
By definition of $|n_{ba}\rangle$, 
it follows that $|n_{ba}\rangle$ is an eigenstate of $\beta_{ab}(\{G_{a'b'}\}) n_i[X^-_{ba}X^+_{ab}]\widehat{L}^i$ 
with maximal, and in particular positive, eigenvalue, whence  $P_{ba}(\{G_{a'b'}\}) | n_{ba} \rangle =   | n_{ba} \rangle$,
for all $a<b$.  
But this in turn implies that $\exp(\Re S_{\gamma < 1}) = \exp(\Re S^{EPRL}_{\gamma < 1})= 1$, 
where the last equality follows from orientation,
as proven in section V.A.2 of \cite{bdfgh2009}.

\noindent($\Rightarrow$) Suppose $\exp(\Re S_{\gamma < 1}) = 1$.  
Then both inequalities in (\ref{gl1ineq}) are equalities. In particular, this implies
\begin{equation}
\label{projid_gl1}
P_{ba}(\{G_{a'b'}\}) |n_{ba} \rangle = |n_{ba}\rangle
\end{equation}
\end{subequations}
for all $a<b$, which in turn implies that $\exp(\Re S^{EPRL}_{\gamma <1}) = \exp(\Re S_{\gamma <1}) =1$,
which, from section V.A.2 in \cite{bdfgh2009}, implies orientation. 
As argued above, this implies that, for all $a \neq b$, 
$n_{ba}^i = \xi_{ba} n[X_{ba}^- X_{ab}^+]^i$ for some $\xi_{ba} = \pm 1$.
From the definition of $|n_{ba}\rangle$, one then has 
$\beta_{ab}(\{G_{a'b'}\}) n[X_{ba}^-X_{ab}^+]_i \hat{L}^i |n_{ba}\rangle = 
\beta_{ab}(\{G_{a'b'}\}) \xi_{ba} k_{ab} |n_{ba}\rangle$.
Equation (\ref{projid_gl1}) then implies the eigenvalue in the foregoing equation is positive for all $a<b$, 
so that $\xi_{ba} = \beta_{ab}(\{G_{a'b'}\})$, 
whence 
$\beta_{ab}(\{G_{a'b'}\}) n[X^-_{ba}X^+_{ab}] \cdot n_{ba} = 1 \ge 0$,
proving proper orientation.
%
%
\finishproof}

\begin{lemma}
\label{gl1stat}
Let boundary data $\{k_{ab}, n_{ab}\}$ be given, and suppose
$\{G_a\}$ is a maximal point
of $S_{\gamma < 1}$. Then it is also a stationary point of $S_{\gamma < 1}$
iff closure is additionally satisfied.
\end{lemma}
{\startproof
If $\delta$ is any variation of the group elements $G_a$, from (\ref{gl1act}),
(\ref{gl1EPRLact}) and the fact that $\{G_a\}$ is maximal, one  has
\begin{equation}
\label{gl1stat_a}
\delta \exp(S_{\gamma < 1}) = \delta \exp(S_{\gamma < 1}^\EPRL)
+
 \prod_{a<b}
\langle J \simpiota{ab} n_{ab}, \quad
\rho(G_{ab}) \simpiota{ab}
\left(\delta P_{ba}(\{G_{a'b'}\})\right) n_{ba} \rangle.
\end{equation}
From lemma \ref{iotalemm}.c,
\begin{equation}
\label{Pintertwine}
P_{ba}(\{G_{a'b'}\}) \circ
\simpiota{ab} =
\simpiota{ab} \circ P_{ba}(\{G_{a'b'}\}).
\end{equation}
Taking the variation of both sides and using the result in (\ref{gl1stat_a}),
\begin{equation}
\label{gl1stat_b}
\delta \exp(S_{\gamma < 1}) =
\delta \exp(S_{\gamma < 1}^\EPRL)
+
 \prod_{a<b}
\langle J \simpiota{ab} n_{ab}, \quad
\rho(G_{ab})  \left(\delta P_{ba}(\{G_{a'b'}\})\right) \simpiota{ab}  n_{ba} \rangle .
\end{equation}
From lemma \ref{gl1max}, as $\{G_a\}$ is a maximal point, orientation and proper orientation are satisfied.
From orientation,
\begin{eqnarray}
\nonumber
\rho(G_{ba}) J \simpiota{ab} |n_{ab}\rangle
&\propto&
\rho(G_{ba}) J \njm{n_{ab}}{\simpj^-_{ab}, \simpj^+_{ab},\rotk_{ab}}{\rotk_{ab}} 
\; \propto \; \rho(G_{ba})
\njm{-n_{ab}}{\simpj^-_{ab}, \simpj^+_{ab},\rotk_{ab}}{\rotk_{ab}} \\
\nonumber
&\propto& \rho(G_{ba})
\left[\njm{-n_{ab}}{\simpj^-_{ab}}{ \simpj^-_{ab}} \otimes
\njm{-n_{ab}}{\simpj^+_{ab}}{ \simpj^+_{ab}}\right] \\
\nonumber
&\propto&
\njm{- X_{ba}^- \triangleright n_{ab}}{\simpj^-_{ab}}{ \simpj^-_{ab}}  \otimes
\njm{- X_{ba}^+ \triangleright n_{ab}}{\simpj^+_{ab}}{ \simpj^+_{ab}} \\
\nonumber
&\propto&
\njm{n_{ba}}{\simpj^-_{ab}}{ \simpj^-_{ab}}  \otimes
\njm{n_{ba}}{\simpj^+_{ab}}{ \simpj^+_{ab}}
\propto \njm{n_{ba}}{\simpj^-_{ab}, \simpj^+_{ab}, \rotk_{ab}}{\rotk_{ab}} \\
\label{gl1stat_c}
&\propto& \simpiota{ab} |n_{ba}\rangle
\end{eqnarray}
where lemma \ref{iotalemm}.b was used in line 1 and  $\rotk_{ab}=\simpj^-_{ab} + \simpj^+_{ab}$,
was used in lines 2 and 4. Furthermore, from orientation and proper orientation, 
by the same argument used in lemma \ref{gl1max}, we have that 
$|n_{ba}\rangle_{k_{ab}}$ is an eigenstate of $P_{ba}(\{G_{a'b'}\})$ with eigenvalue $1$, so that, 
by equation (\ref{Pintertwine}), $\simpiota{ab}|n_{ba}\rangle_{k_{ab}}$ is also an eigenstate of $P_{ba}(\{G_{a'b'}\})$
with eigenvalue $1$. This, together with (\ref{gl1stat_c}), via corollary \ref{projvarcor} in appendix \ref{projvarapp},
implies that the second term in (\ref{gl1stat_b})
is zero.  As proven in \cite{bdfgh2009}, using the fact that orientation is satisfied,
the remaining term in (\ref{gl1stat_b}) is zero iff closure is satisfied.
\finishproof}

\begin{theorem}
\label{gl1thm}
Given boundary data $\{k_{ab}, n_{ab}\}$, $\{G_a\}$ is a
critical point of $S_{\gamma < 1}$ iff closure, orientation, and proper orientation
are satisfied.
\end{theorem}
{\startproof

($\Rightarrow$)
Suppose $\{G_a\}$ is a
critical point of $S_{\gamma < 1}$.  Then lemma \ref{gl1max} implies that
orientation and proper orientation are satisfied, and  lemma \ref{gl1stat}
implies that closure is satisfied.

($\Leftarrow$)
Suppose closure, orientation, and proper orientation
are satisfied. Then by lemma \ref{gl1max},  $\{G_a\}$ is a maximal point
of $S_{\gamma <1}$, and by lemma \ref{gl1stat} it is a stationary point
of $S_{\gamma < 1}$.
\finishproof}

\subsubsection{The case $\gamma > 1$}

For this case, we derive from scratch an expression for the proper vertex
analogous to (18) and (19) in \cite{bdfgh2009}.  In doing this, we use the spinorial form of the
irreps of $SU(2)$.
Let $A,B, C, \cdots = 0,1$ denote spinor indices.
The carrying space $V_j$ can then be realized as the
space of symmetric  spinors of rank $2j$ (see, for example,
%
%
\cite{rovelli2004}).
Let  $n^A$ denote the spinor corresponding to the coherent state $|n\rangle_{\half}$.
As in \cite{bdfgh2009, cf2008a}, the key property of coherent states we use is that, in
their spinorial form, the higher spin coherent states are given by
 \begin{equation}
 (|n\rangle_j )^{A_1 \cdots A_{2j}} = n^{A_1} \cdots n^{A_{2j}} .
 \end{equation}
From the relation (\ref{skrelation}) between $\rotk$ and $\simpj^+,\simpj^-$ for a given
triangle, one deduces for $\gamma > 1$ that
$\simpj^+ = \simpj^- + \rotk$.
For this case, the explicit expression for $\simpiota{}$ in terms of symmetric spinors is given in equations (A.12) and (A.13) of \cite{rovelli2004}\footnote{In
(A.13) of \cite{rovelli2004}, symmetrization over the $A$ group, $B$ group, and $C$ group of indices
was forgotten but was clear from the context.
}.
Let $v^{A_1 \cdots A_{2\rotk}} \in V_\rotk$ be given.
For $\gamma > 1$, one has
%
%
\begin{equation}
\label{i_ggreater1}
\simpiota{}(v)^{A_{1}\cdots A_{2\simpj^+} B_{1} \cdots B_{2\simpj^-}}
= v^{(A_{1}\cdots A_{2\rotk}} \epsilon^{A_{2\rotk+1}| B_{1}|} \cdots \epsilon^{A_{2\simpj^+}) B_{2\simpj^-}}
\end{equation}
where the symmetrization is over the $A$ indices only.
In order
to impose the symmetrization over the $A$ indices, similar to \cite{bdfgh2009},
on the left of each $\simpiota{}$, acting
in the self-dual part of the codomain, we insert a resolution of the identity
on $V_{\simpj^+}$ into coherent states:
\begin{equation}
d_{\simpj^+}\int \dif m |m\rangle_{s^+} {}_{s^+}\langle m|= I_{\simpj^+}
\end{equation}
where $\dif m$ is the measure on the metric 2-sphere normalized to unit area, and
$d_\simpj:= 2\simpj+1$.
In spinorial notation
\begin{equation}
\label{spinres}
d_{\simpj^+}\int \dif m \;  m^{A_1} \cdots m^{A_{2\simpj^+}}
m^\dagger_{B_1} \cdots m^\dagger_{B_{2\simpj^+}} =
\delta^{(A_1}_{B_1} \cdots \delta^{A_{2\simpj^+})}_{B_{2\simpj^+}} .
\end{equation}
%
where $m^\dagger_A := ({}_{\half}\langle m |)_A$ .
Starting from equation (\ref{propeprl}) with
$\psi_{ab} = |n_{ab}\rangle_{k_{ab}} = n_{ab}^{A_1} \cdots n_{ab}^{A_{2k_{ab}}}$, writing out all spinor
 indices explicitly, we insert two resolutions of the identity (\ref{spinres}) into each face factor
 in (\ref{propeprl}), one after each $\simpiota{}$.
Denote the integration variables $m_{ab}$ and $m_{ba}$  respectively for the left and right
insertions. 
Writing out the $\epsilon$-inner product
in terms of alternating tensors $\epsilon_{AB}$, using $m^\dagger_A = -\epsilon_{AB} (J m)^B$,
simplifying, and then writing the final expression again in terms of Hermitian inner products,
one obtains
\begin{equation}
\label{vertex_gg1}
A^\prop_v = \int \prod_a \dif G_a \left(\prod_{a<b}(-1)^{2\simpj^-_{ab}} d^2_{\simpj^+_{ab}} 
\dif m_{ab} \dif m_{ba}\right) \exp(S_{\gamma > 1})
\end{equation}
where
\begin{eqnarray}
\nonumber
\exp(S_{\gamma > 1}) &=&
\prod_{a<b}
{}_{\rotk_{ab}}\langle m_{ab} | n_{ab} \rangle_{\rotk_{ab}}
{}_{\simpj^+_{ab}}\langle Jm_{ab} | \rho(X_{ab}^+)
| m_{ba} \rangle_{\simpj^+_{ab}} \\
\label{gg1act}
&& \dummy \qquad
{}_{\rotk_{ab}}\langle m_{ba} | P_{ba}(\{G_{a'b'}\}) | n_{ba} \rangle_{\rotk_{ab}}
\overline{{}_{\simpj^-_{ab}}\langle Jm_{ab} | \rho(X_{ab}^-) | m_{ba} \rangle_{\simpj^-_{ab}}} .
\end{eqnarray}
%
%
%
%
Recall from \cite{bdfgh2009} that the action for $\gamma >1$ for the original EPRL model
is\footnote{The
coherent state $|m_{ab}\rangle$ used here is related to the
corresponding coherent state used in \cite{bdfgh2009} by the action of $J$.
%
%
}
\begin{eqnarray}
\nonumber
\exp(S^\EPRL_{\gamma > 1})
&=&
\prod_{a<b}
{}_{\rotk_{ab}}\langle m_{ab} | n_{ab} \rangle_{\rotk_{ab}}
{}_{\simpj^+_{ab}}\langle Jm_{ab} | \rho(X_{ab}^+)
| m_{ba} \rangle_{\simpj^+_{ab}} \\
\label{gg1EPRLact}
&& \dummy \qquad
{}_{\rotk_{ab}}\langle m_{ba} | n_{ba} \rangle_{\rotk_{ab}}
\overline{{}_{\simpj^-_{ab}}\langle Jm_{ab} | \rho(X_{ab}^-) | m_{ba} \rangle_{\simpj^-_{ab}}} .
\end{eqnarray}

\begin{lemma}
\label{gg1max}
Given boundary data $\{k_{ab}, n_{ab}\}$, $\{G_a, m_{ab}\}$ is a maximal point
of $S_{\gamma > 1}$ iff orientation and
proper orientation are satisfied and $m_{ab} = n_{ab}$ for all $a \neq b$.
\end{lemma}
{\startproof
From (\ref{gg1act}),
%
%
\begin{subequations}
\begin{eqnarray}
\nonumber
\exp(\Re S_{\gamma > 1}) &=& |\exp (S_{\gamma > 1})|  \\
\nonumber
&=& \prod_{a<b} \Big| {}_{k_{ab}}\!\langle m_{ab}| n_{ab}\rangle_{k_{ab}}\Big| \,
 \Big| {}_{s^+_{ab}}\!\langle J m_{ab}| \rho(X^+_{ab}) |m_{ba}\rangle_{s^+_{ab}}\Big| \,
  \Big| {}_{k_{ab}}\!\langle m_{ba}| P_{ba}(\{G_{a'b'}\})|n_{ba}\rangle_{k_{ab}}\Big| \cdot  \\
\nonumber
&& \hspace{1.5cm}
   \cdot \Big| {}_{s^-_{ab}}\!\langle J m_{ab}| \rho(X^-_{ab}) |m_{ba}\rangle_{s^-_{ab}}\Big| \\
\label{gg1ineq}
&\le& \prod_{a<b} || P_{ba}(\{G_{a'b'}\}) |n_{ba}\rangle_{k_{ab}} || \le 1
\end{eqnarray}
where the Cauchy-Schwarz inequality,
the fact that $J$ and $\rho(X^\pm_{ab})$ are norm preserving,
and $|| \, |n_{ab}\rangle||=|| \, |m_{ab}\rangle||=1$  have been used in the first inequality, and  
$|| \, |n_{ba}\rangle||=1$ has been used in the last inequality.

We now proceed to prove that $\exp(\Re S_{\gamma > 1})=1$ 
iff orientation and proper orientation are satisfied and $m_{ab} = n_{ab}$ for all $a \neq b$.

\noindent($\Leftarrow$) Suppose orientation and proper orientation are satisfied and $m_{ab}=n_{ab}$ for all $a\neq b$.  
By the same argument used in lemma \ref{gl1max}, it follows that $P_{ba}(\{G_{a'b'}\}) | n_{ba} \rangle =   | n_{ba} \rangle$
for all $a<b$.  
But this in turn implies that $\exp(\Re S_{\gamma > 1}) = \exp(\Re S^{EPRL}_{\gamma > 1})= 1$, 
where the last equality follows from orientation and $m_{ab} = n_{ab}$,
as proven in section V.A.2 of \cite{bdfgh2009}.

\noindent($\Rightarrow$) Suppose $\exp(\Re S_{\gamma > 1}) = 1$.  Then both 
inequalities in (\ref{gg1ineq}) are equalities. In particular it follows
\begin{equation}
\label{projid_gg1}
P_{ba}(\{G_{a'b'}\}) |n_{ba} \rangle = |n_{ba}\rangle
\end{equation}
\end{subequations}
for all $a<b$, which in turn implies that $\exp(\Re S^{EPRL}_{\gamma >1}) = \exp(\Re S_{\gamma >1}) =1$,
which, from section V.A.2 in \cite{bdfgh2009}, implies orientation and $m_{ab} = n_{ab}$.  Furthermore, 
by the same argument used in lemma \ref{gl1max},  (\ref{projid_gg1}) also implies proper orientation.

\finishproof}

\begin{lemma}
\label{gg1stat}
Let boundary data $\{k_{ab}, n_{ab}\}$ be given, and suppose
$\{G_a, m_{ab}\}$ is a maximal point
of $S_{\gamma > 1}$. Then it is also a stationary point of $S_{\gamma > 1}$
iff closure is additionally satisfied.
\end{lemma}
{\startproof
If $\delta$ is any variation of $G_a$ and  $m_{ab}$, from (\ref{gg1act})
and (\ref{gg1EPRLact}) one has
\begin{eqnarray}
\nonumber
\delta \exp(S_{\gamma > 1}) &=& \delta \exp(S_{\gamma > 1}^\EPRL)
+
\prod_{a<b}
{}_{\rotk_{ab}}\langle m_{ab} | n_{ab} \rangle_{\rotk_{ab}}
{}_{\simpj^+_{ab}}\langle Jm_{ab} | \rho(X_{ab}^+)
| m_{ba} \rangle_{\simpj^+_{ab}} \\
\label{gg1stat_a}
&& \dummy \qquad
{}_{\rotk_{ab}}\langle m_{ba} | (\delta P_{ba}(\{G_{a'b'}\})) | n_{ba} \rangle_{\rotk_{ab}}
\overline{{}_{\simpj^-_{ab}}\langle Jm_{ab} | \rho(X_{ab}^-) | m_{ba} \rangle_{\simpj^-_{ab}}} .
\end{eqnarray}
Because $\{G_a, m_{ab}\}$ is a maximal point,
from lemma \ref{gg1max}, orientation and proper orientation are satisfied,
and $m_{ab} = n_{ab}$ for all $a\neq b$.
It follows that
%
%
$|n_{ba}\rangle_{k_{ab}} = |m_{ba}\rangle_{k_{ab}}$ 
and, by the same argument used in lemma \ref{gl1max}, 
both are eigenstates of $P_{ba}(\{G_{a'b'}\})$ with eigenvalue 1, so that
by corollary \ref{projvarcor} in appendix \ref{leftapp}, the second
term above is zero.
As proven in \cite{bdfgh2009}, because orientation
is satisfied and $m_{ab}=n_{ab}$ for all $a\neq b$, it follows that
the remaining term in (\ref{gg1stat_a}) is zero iff closure is satisfied.
\finishproof}

\begin{theorem}
\label{gg1thm}
Given boundary data $\{k_{ab}, n_{ab}\}$, $\{G_a, m_{ab}\}$ is a
critical point of $S_{\gamma > 1}$ iff closure, orientation, and proper orientation
are satisfied, and $m_{ab}=n_{ab}$ for all $a \neq b$.
\end{theorem}
{\startproof

($\Rightarrow$)
Suppose $\{G_a, m_{ab}\}$ is a
critical point of $S_{\gamma > 1}$.  Then lemma \ref{gg1max} implies that
orientation and proper orientation are satisfied and $m_{ab} = n_{ab}$
for all $a \neq b$, and  lemma \ref{gg1stat}
implies that closure is satisfied.

($\Leftarrow$)
Suppose closure, orientation, and proper orientation
are satisfied and $m_{ab} = n_{ab}$ for all $a \neq b$.
Then by lemma \ref{gg1max},  $\{G_a, m_{ab}\}$ is a maximal point
of $S_{\gamma >1}$, and by lemma \ref{gg1stat} it is a stationary point
of $S_{\gamma > 1}$.
\finishproof}

Thus, though in the $\gamma >1$ case one has an extra set of variables
$\{m_{ab}\}$, these are restricted to be equal to $\{n_{ab}\}$ by the critical point equations,
allowing one to treat the $\gamma < 1$ and $\gamma > 1$ cases in a unified way. The remaining
critical point conditions on $\{G_a, n_{ab}\}$ (given
in theorem \ref{gl1thm}) have a symmetry: 
if $\{G_a\}$ form a solution, 
then so does the set of group elements $\{\tilde{G}_a = (\tilde{X}^-_a, \tilde{X}^+_a)\}$ with
\begin{equation}
\label{critsym}
\tilde{X}^\pm_a = \epsilon^\pm_a \ggk^\pm X^\pm_a
\end{equation}
for any $(\ggk^-, \ggk^+) \in Spin(4)$ and any set of ten signs $\epsilon^\pm_a$.  This transformation
is also a symmetry of the actions (\ref{gl1act}) and (\ref{gg1act}). If two solutions
$\{G_a\}$, $\{\tilde{G}_a\}$ are related by such a symmetry transformation, we call them
\textit{equivalent} and write $\{G_a\} \sim \{\tilde{G}_a\}$.
%
%

\subsection{Proof of the asymptotic formula}

Using the above results, we proceed to prove theorem \ref{prop_asym_thm}.

Before getting into the details of the proof, we summarize its general structure.
As already mentioned, the critical point equations
for the proper vertex integrals (\ref{gl1vert}) and (\ref{vertex_gg1})
have a set of symmetries (\ref{critsym}), of
which the global $Spin(4)$ symmetry is the only continuous one.  In order to apply the stationary
phase method to calculate the asymptotics, the critical points must be isolated, and hence this
continuous symmetry must be removed.  As in \cite{bdfgh2009}, we do this by performing
the change of variables $\tilde{G}_a:= (G_0)^{-1}G_a$ for $a=1,\dots 4$.
Then $G_{0}$ no longer appears in the integrand, so that the $G_0$ integral
drops out. Upon removing the tilde labels, the remaining integrand is the same as the original integrand
except with $G_0$ replaced by the
identity.  In what follows $G_a = (X^-_a, X^+_a)$ shall denote these ``gauge-fixed'' group elements,
with $G_0 \equiv \mathrm{id}$,
in terms of which the continuous symmetry has been removed.
%
%

The proof then has two steps, the first of which has already been done in theorems \ref{gl1thm}
and \ref{gg1thm} above: (1.) prove that the critical points of proper EPRL are
precisely the subset of critical points of original EPRL at which
proper orientation is satisfied. (2.)
prove that, given a set of $SU(2)$ boundary data $\{k_{ab}, n_{ab}\}$,
the critical points of original EPRL at which
proper orientation is satisfied are all
equivalent and are precisely the critical points which give rise to the asymptotic term
(\ref{prop_asym}) in the original EPRL asymptotics \cite{bdfgh2009}.
Because proper orientation is satisfied, the projector $P_{ba}$ will act as the identity, and
the value of the proper EPRL action at these critical points will be the same as
the value of the original
EPRL action at these points, yielding precisely the asymptotic behavior (\ref{prop_asym})
claimed.

Let us begin by reviewing the results from theorems \ref{gl1thm} and \ref{gg1thm}.
The critical point equations for $\gamma < 1$ and $\gamma > 1$ are equivalent: the only difference is
that for $\gamma > 1$ one integrates over extra variables, $m_{ab}$,
which, however, come with the critical point equations $m_{ab} = n_{ab}$, eliminating them.
This allows us to effectively consider both the $\gamma < 1$ case and $\gamma > 1$ case simultaneously in the following.
%
%
As given in theorems \ref{gl1thm} and \ref{gg1thm},
the remaining critical point equations are
\begin{equation}
\label{orient_constr}
X_a^{\pm} \triangleright n_{ab} = - X_b^{\pm} \triangleright n_{ba}
\end{equation}
and
\begin{equation}
\label{prop_constr}
\beta_{ab} \tr(\tau_i X^-_{ab} X^+_{ba})n^i_{ab} > 0
\end{equation}
for all $a < b$.
The first of these, (\ref{orient_constr}), is of the same form for both $\{X_a^+\}$
and $\{X_a^-\}$:
\begin{equation}
\label{ueqn}
U_a \triangleright n_{ab} = - U_b \triangleright n_{ba} .
\end{equation}
One therefore proceeds by finding the solutions $\{U_a\}$ to (\ref{ueqn}) for a given
set of $SU(2)$ boundary data $\{k_{ab},n_{ab}\}$, and then from these one constructs the solutions $\{G_a\}$
to (\ref{orient_constr}), and then one checks which among these, if any, solves (\ref{prop_constr})
in order to determine the critical points of the vertex integral.

The solutions to (\ref{ueqn}) have already been analyzed by \cite{bdfgh2009}.
To use the results in this reference, one needs
the notion of a \textit{vector geometry}:
A set of boundary data $\{k_{ab}, n_{ab}\}$ is called a \textit{vector geometry}
if it satisfies closure and there exists $\{h_a\} \subset SO(3)$ such that
$(h_a \triangleright n_{ab})^i =  - (h_b \triangleright n_{ba})^i$ for all $a \neq b$.
The authors of \cite{bdfgh2009} then proceed by
considering separately the three cases
 in which the boundary data
(i.) does not define a vector geometry (ii.) defines a vector geometry which is, however, not a
nondegenerate 4-simplex geometry, and (iii.) defines a nondegenerate 4-simplex geometry.
We use this same division and consider each of these three cases in turn.

\noindent\textit{Case (i.): Not a vector geometry.}

In this case, as proven in \cite{bdfgh2009}, there are no solutions to (\ref{ueqn}) and hence
no solutions to (\ref{orient_constr}), and hence no critical points.  The
vertex integral therefore decays exponentially with $\lambda$.

\noindent\textit{Case (ii.): A vector geometry, but no nondegenerate 4-simplex geometry.}

In this case, as proven in \cite{bdfgh2009}, there is exactly one solution to (\ref{ueqn}),
up to the equivalence (\ref{eqrel}).  The only solution to (\ref{orient_constr}) is therefore
$(X_a^-, X_a^+) = (U_a,\epsilon_a Y U_a)$. But then $X^-_{ba} X^+_{ab} = \pm I$,
so that this solution fails to satisfy condition (\ref{prop_constr}), so that there are no
critical points.  The vertex integral therefore decays exponentially with $\lambda$.

\noindent\textit{Case (iii.): A nondegenerate 4-simplex geometry.}

In this case, as proven in \cite{bdfgh2009},
(\ref{ueqn}) has two inequivalent solutions $\{U_a^1\}$ and $\{U_a^2\}$, so that there are four
inequivalent solutions to (\ref{orient_constr}):
$(X_a^-, X_a^+) = (U_a^1, U_a^1), (U_a^2, U_a^2), (U_a^1, U_a^2), (U_a^2, U_a^1)$.
%
%
Neither $(U_a^1, U_a^1)$ nor $(U_a^2, U_a^2)$,  nor any solution equivalent to these, satisfies
(\ref{prop_constr}), again because $X^-_{ba} X^+_{ab} = \pm I$.
It remains only to consider the solutions
\begin{eqnarray}
\nonumber
(\Xdec{1}^-_a,\Xdec{1}^+_a) &=& (U^1_a, U^2_a) \\
\label{lastsolns}
(\Xdec{2}^-_a,\Xdec{2}^+_a) &=& (U^2_a, U^1_a) .
\end{eqnarray}
Because $\Xdec{1}^-_{ab} \Xdec{1}^+_{ba} = \left( \Xdec{2}^-_{ab} \Xdec{2}^+_{ba}\right)^{-1}$, the proper axes
$n[\Xdec{1}^-_{ab} \Xdec{1}^+_{ba}]^i$,
$n[\Xdec{2}^-_{ab} \Xdec{2}^+_{ba}]^i$ defined in
(\ref{SUtwoexp}) are equal and opposite, so that
\begin{equation}
\label{signopp}
\tr (\tau_i \Xdec{1}^-_{ab} \Xdec{1}^+_{ba})
= - \tr (\tau_i \Xdec{2}^-_{ab} \Xdec{2}^+_{ba}).
\end{equation}
From this one deduces
\begin{equation}
\label{betasame}
\beta_{ab}(\{\Gdec{1}_{a'b'}\}) = \beta_{ab}(\{\Gdec{2}_{a'b'}\})
\end{equation}
which gives us
\begin{equation}
\label{muopp}
\beta_{ab}(\{\Gdec{1}_{a'b'}\})\tr (\tau_i \Xdec{1}^-_{ab} \Xdec{1}^+_{ba}) n_{ba}^i
= - \beta_{ab}(\{\Gdec{2}_{a'b'}\})  \tr (\tau_i \Xdec{2}^-_{ab}
\Xdec{2}^+_{ba}) n_{ba}^i
\end{equation}
for all $a\neq b$.
Because $\{U_a^1\} \not\sim \{U_a^2\}$, we have
$\{\Xdec{1}_a^+\} \not\sim \{\Xdec{1}_a^-\}$ and
$\{\Xdec{2}_a^+\} \not\sim \{\Xdec{2}_a^-\}$, so that both
$\{\Gdec{1}_a\}$ and $\{\Gdec{2}_a\}$ satisfy the hypotheses
%
%
of lemma \ref{mulemm}, implying that
%
%
neither side of (\ref{muopp}) is zero.  It follows that exactly
one of
$\beta_{ab}(\{\Gdec{\alpha}_{a'b'}\})\tr (\tau_i \Xdec{\alpha}^-_{ab}
\Xdec{\alpha}^+_{ba}) n_{ba}^i$, $\alpha = 1,2$,
is positive, so that exactly one of
$\{\Gdec{1}_a\}$, $\{\Gdec{2}_a\}$  satisfies
proper orientation and so is a critical point.
Furthermore, at this one critical point, by theorem \ref{classcon}, the $\mu$ arising in the reconstruction theorem
(theorem \ref{reconth} here and theorem 3 in \cite{bdfgh2009}) is $1$.
Because, at this point, the value of the action (\ref{gl1act}) (respectively (\ref{gg1act}))
for the proper vertex is equal to the
value of the action (\ref{gl1EPRLact})(respectively (\ref{gg1EPRLact}))
for the original vertex, from the analysis of \cite{bdfgh2009},
this one critical point gives rise  to precisely the desired
asymptotics stated in theorem \ref{prop_asym_thm}.

\section{Conclusions}

The original EPRL model, as shown and emphasized in \cite{engle2011, engle2011err},
due to the fact that it is based on the linear simplicity constraints, necessarily mixes three
of what we call Plebanski sectors as well as two dynamically determined orientations.
This mixing of sectors was  identified as the precise reason
for the multiplicity of terms in the asymptotics of the EPRL vertex calculated in \cite{bdfgh2009}.
Furthermore, when multiple 4-simplices are considered, asymptotic analysis thus far \cite{mp2011, hz2011}
indicates that critical configurations contribute in which these sectors can \textit{vary locally}
from 4-simplex to 4-simplex.
The asymptotic amplitude for such configurations is the
exponential of $i$ times an action which
is not Regge, but rather a sort of `generalized Regge action'.
%
%
The stationary points of this `generalized' action are not in general solutions to the Regge equations
of motion, and thus one has sectors in the semiclassical limit which do not represent general relativity.

In this paper, a solution to this problem is found.
We began by deriving a classical discrete condition that isolates the sector corresponding to the first term in the asymptotics
--- what we have called the Einstein-Hilbert sector, 
in which the BF action is equal to the Einstein-Hilbert action including sign.  Equivalently, this is the sector in which
the sign of the Plebanski sector (II$\pm$) matches the sign of the dynamical orientation.
By appropriately quantizing this condition and using it to modify
the EPRL vertex amplitude, we have constructed what we call the \textit{proper} EPRL vertex amplitude.
This vertex amplitude continues to be a function of $SU(2)$ spin-network data, so that it may continue
to be used to define dynamics for LQG.  We have shown that
the proper vertex is $SU(2)$ gauge invariant and is linear in the boundary state,
as required to ensure that the final transition amplitude is linear in the initial state and antilinear in the final state.
It is furthermore $Spin(4)$-invariant in the sense that, similar to the original EPRL model \cite{rs2010a}, 
it is independent of the choice of extra structures used in its definition which seem to break $Spin(4)$ symmetry.
Finally, it has the correct asymptotics with the \textit{single} term consisting in the exponential of $i$ times the Regge action.

Two interesting further research directions would be (1.) to justify the Lorentzian signature generalization
given in equation (\ref{lorprop}), via a quantization of the Lorentzian Einstein-Hilbert sector, and to verify that 
(\ref{lorprop}) also has the desired single-termed semiclassical limit and 
(2.) to generalize the present work to the amplitude for an arbitrary 4-cell, which might be used in a spin-foam model involving
arbitrary cell-complexes, similar to the generalization \cite{kkl2009} of Kami\'nski, Kisielowski, and Lewandowski.
The first of these tasks should be straightforward.  The second, however,
seems to require a new way of thinking about the discrete constraint (\ref{EHineq}) used to isolate the Einstein-Hilbert sector.
For, the $\beta_{ab}$ sign factor involved in this condition uses in a central way the fact that there are 5 tetrahedra in
each 4-simplex.

Lastly, it is important to understand if and how the graviton propagator calculations  \cite{rovelli2005, ar2007, ar2007a, bmp2009b, bd2011},
and  spin-foam cosmology calculations  \cite{
rv2008, brv2010, vidotto2011}
will change if the presently proposed proper vertex is used in place of the original EPRL
vertex.  In the case of the graviton propagator, only the leading order term in the vertex expansion
has thus far been calculated \cite{bmp2009b}.  To this order, only one 4-simplex is involved, and one does not expect
the use of the proper vertex to change the results, because the coherent boundary state used in this work already
suppresses all but the one desired term (\ref{prop_asym}) in the asymptotics.  However, higher order terms in the propagator may very well be affected by the use of the proper vertex.  We leave this and similar such questions for future investigations.

\section*{Acknowledgements}
The author thanks Christopher Beetle for invaluable discussions and for pointing
out a simpler proof for theorem \ref{geombivth},
Carlo Rovelli and Antonia Zipfel for remarks on a prior draft,
Alejandro Perez and Abhay Ashtekar for encouraging the author to finish this work,
and the quantum gravity institute at the University of Erlangen-Nuremburg for an invitation 
to give a seminar on this topic, which led to the correction of an important sign error. This work was supported in part by the
NSF through grant  PHY-1237510 and by
the National Aeronautics and Space Administration
through the University of Central Florida's NASA-Florida Space Grant Consortium.

\appendix

\section{Well-definedness of orientation and Plebanski sectors}
\label{defapp}

As in section \ref{clsect} in the main text, we let $\mathcal{B}_{ab}$ denote the bivectors $G_a \triangleright B_{ab}$
in the 4-simplex frame. Throughout this appendix we assume that
$G_a$ and $B_{ab}$ satisfy closure, orientation, and linear simplicity, implying
corresponding restrictions on $\mathcal{B}_{ab}$.\footnote{The consequences of linear simplicity
will only be used in the final lemma.}
As mentioned in section
\ref{clsect}, for each choice of flat connection $\partial_\mu$ adapted to $S$,
%
%
the discrete variables $\{\mathcal{B}_{ab}\}$ determine a unique continuum two form $B_{\mu\nu}(\{\mathcal{B}_{ab}\}, \partial)$
via the conditions $\partial_\sigma B_{\mu\nu}(\{\mathcal{B}_{ab}\}, \partial) = 0$ and
$\int_{\Delta_{ab}(S)} B(\{\mathcal{B}_{cd}\}, \partial) = \mathcal{B}_{ab}$.
This continuum two form in turn determines a dynamical orientation of $S$, as well as determining one of
three Plebanski sectors, represented respectively by the functions
$\omega(B_{\mu\nu})$ and $\nu(B_{\mu\nu})$, defined in section \ref{clframe},
taking values in $\{0,1,-1\}$.
We here prove that the orientation and Plebanski sector of $B_{\mu\nu}(\{\mathcal{B}_{ab}\}, \partial)$ are
independent of the choice of $\partial_\mu$ adapted to $S$.  (In the paper
\cite{engle2011, engle2011err}, a slightly different but equivalent
way of reconstructing $B_{\mu\nu}$ was used.  The well-definedness of
orientation and Plebanski sector of $B_{\mu\nu}$ as reconstructed there
was proven in that paper. We here prove it anew for the new present reconstruction of $B_{\mu\nu}$, for completeness.)
%
%

In the following, we denote the vertex of $S$ opposite each tetrahedron $a \in \{0,\dots,4\}$ by $p_a$.  We also use
the term `face' in the general sense of any lower dimensional simplex which forms part of the boundary of a higher
dimensional simplex. Bold lower case Greek letters, $\boldsymbol\mu, \boldsymbol\nu=0,1,2,3$, shall be used to label
different coordinates of a given coordinate system.

\begin{lemma}
\label{coordlem}
Given a flat connection $\partial_\mu$ adapted to $S$, there exists a unique coordinate system $x^{\boldsymbol\mu}$
such that (1.) $\partial_\mu$ is the associated coordinate derivative operator and
(2.) the values of the coordinates at the five vertices
of $S$, $p_0$, $p_1$, $p_2$, $p_3$, $p_4$, are respectively given by
\begin{equation}
\label{canvert}
(1,0,0,0), (0,1,0,0),
(0,0,1,0), (0,0,0,1), \text{ and }(0,0,0,0).
\end{equation}
Furthermore, the range of possible values of the 4-tuple of coordinates
$(x^{\boldsymbol\mu})$ coincides precisely with the linear 4-simplex in $\R^4$ with these five vertices, which we refer to as the
`canonical 4-simplex' in $\R^4$.
We call $\{x^{\boldsymbol\mu}\}$ `the coordinate system determined by $\partial_\mu$'.
\end{lemma}
{\startproof
For each $\boldsymbol\mu = 0,1,2,3$, let $V_{\boldsymbol\mu}^\mu$ denote a vector in $T_{p_4}S$
tangent to the edge $\overline{p_4 p_{\boldsymbol\mu}}$, pointing away from $p_4$.  This vector is unique up to scaling
by a positive number.  Fix this scaling freedom by first parallel transporting $V_{\boldsymbol\mu}^\mu$ along the edge
$\overline{p_4 p_{\boldsymbol\mu}}$, and then requiring that the affine length of  $\overline{p_4 p_{\boldsymbol\mu}}$
with respect to $V_{\boldsymbol\mu}^\mu$ be 1.

Because the connection $\partial_\mu$ is flat, one can use $\partial_\mu$
to unambiguously parallel transport $V_{\boldsymbol\mu}^\mu$ to all of $S$, yielding
a vector field $V_{\boldsymbol\mu}^\mu$ on $S$ for each
$\boldsymbol\mu=0,1,2,3$.  Because the vectors $\{V_{\boldsymbol\mu}^\mu(p_4)\}$ at $p_4$ were chosen linearly independent,
the vectors $\{V_{\boldsymbol\mu}^\mu(p)\}$ at each point $p \in S$
form a basis of $T_p S$. Let $\{\lambda_\mu^{\boldsymbol\mu}(p)\}$ denote the basis dual to
$\{V_{\boldsymbol\mu}^\mu(p)\}$ at each $p$.
For each $\boldsymbol \mu$, the resulting one form $\lambda_\mu^{\boldsymbol\mu}$ then
satisfies $\partial_\mu \lambda_\nu^{\boldsymbol\mu} = 0$,
implying $\partial_{[\mu} \lambda_{\nu]}^{\boldsymbol\mu} = 0$; because $S$
is simply connected, this implies that, for each $\boldsymbol\mu$, there exists a function $x^{\boldsymbol\mu}$,
unique up to addition of a constant, such that $\lambda_{\mu}^{\boldsymbol\mu} = \partial_\mu x^{\boldsymbol\mu}$.
Fix this freedom in each $x^{\boldsymbol\mu}$ by requiring $x^{\boldsymbol\mu}(p_4) = 0$.
Because $\lambda_{\mu}^{\boldsymbol\mu} = \partial_\mu x^{\boldsymbol\mu}$ are everywhere linearly independent,
$\{x^{\boldsymbol\mu}\}$ forms a good coordinate system on $S$.
Furthermore, from $V_{\boldsymbol\mu}^\mu \partial_\mu x^{\boldsymbol\nu}
= V_{\boldsymbol\mu}^\mu \lambda_\mu^{\boldsymbol\nu} =
\delta^{\boldsymbol\nu}_{\boldsymbol\mu}$,
one has $V_{\boldsymbol\mu}^\mu = \left(\frac{\partial}{\partial x^{\boldsymbol\mu}}\right)^\mu$.
Because $\partial_\mu$ annihilates $\lambda_\mu^{\boldsymbol\mu} = \partial_\mu x^{\boldsymbol\mu}$
(and $V_{\boldsymbol\mu}^\mu = \left(\frac{\partial}{\partial x^{\boldsymbol\mu}}\right)^\mu$), $\partial_\mu$
is the coordinate derivative operator for $\{x^{\boldsymbol\mu}\}$.
%
%

Consider the differential equation $V_{\boldsymbol\mu}^\mu \partial_\mu x^{\boldsymbol\nu} = \delta^{\boldsymbol\nu}_{\boldsymbol\mu}$
for a given fixed $\boldsymbol\mu$. Because $V_{\boldsymbol\mu}^\mu$
is tangent to the edge $\overline{p_4 p_{\boldsymbol\mu}}$, this equation dictates how to evolve each of the four coordinates
$x^{\boldsymbol\nu}$ along $\overline{p_4 p_{\boldsymbol\mu}}$  from its starting value $x^{\boldsymbol\nu}=0$ at $p_4$,
thereby determining its value at $p_{\boldsymbol\mu}$.
For ${\boldsymbol\nu} \neq {\boldsymbol\mu}$ this implies $x^{\boldsymbol\nu} = 0$ at
$\overline{p_4 p_{\boldsymbol\mu}}$.
For $\boldsymbol\nu = \boldsymbol\mu$, the differential equation simply expresses that $x^{\boldsymbol\mu}$ is an affine coordinate for
$V_{\boldsymbol\mu}^\mu$ along  $\overline{p_4 p_{\boldsymbol\mu}}$, so that, by construction,
$x^{\boldsymbol\mu} = 1$
at $p_{\boldsymbol\mu}$.

Now, the coordinates $x^{\boldsymbol\mu}$ provide an embedding $\Phi$ of $S$ into $\R^4$,
$\Phi: p \mapsto x^{\boldsymbol\mu}(p)$.
By construction the point $p_4$ maps to $(0,0,0,0)$, whereas, as just shown, the points $p_0, p_1, p_2, p_3$
map to the points  $(1,0,0,0)$, $(0,1,0,0)$, $(0,0,1,0)$, $(0,0,0,1)$.
Because $\partial_\mu$ is adapted to $S$, $S$ is the convex hull of its vertices as determined by the affine structure
defined by $\partial_\mu$.  But this affine structure is the same as that defined by the coordinates
$x^{\boldsymbol\mu}$, so that $\Phi[S]$ is the convex hull, in $\R^4$, of the points (\ref{canvert}).
That is, $\Phi[S]$ is the linear 4-simplex in $\R^4$ with vertices (\ref{canvert}).
\finishproof}

For the purposes of the following, the action of a diffeomorphism $\varphi$ on a derivative operator
$\partial_\mu$ is defined by $(\varphi \cdot \partial)_\mu \lambda_\nu
:= (\varphi^{-1})^* \partial_\mu (\varphi^* \lambda_\nu)$.  The resulting action of
$(\varphi \cdot \partial)$ on a general tensor
$t^{\alpha \dots \gamma}{}_{\rho \dots \sigma}$ is then given by
$(\varphi \cdot \partial)_\mu t^{\alpha \dots \gamma}{}_{\rho\dots \sigma}
:=  \varphi \cdot (\partial_\mu (\varphi^{-1} \cdot t^{\alpha \dots \gamma}{}_{\rho \dots \sigma}))$
where $\varphi \cdot$ denotes the left action of $\varphi$ on the tensor in question
(thus, push-forward for contravariant indices and pull-back via $\varphi^{-1}$ for covariant
indices \cite{wald1984}).

\begin{lemma}
\label{ordiffeolem}
Given any two flat connections $\partial_\mu$, $\tilde{\partial}_\mu$ adapted to $S$, there exists
an orientation preserving diffeomorphism $\varphi: S \rightarrow S$ mapping each face of $S$ to itself, and mapping
$\partial_\mu$ to $\tilde{\partial}_\mu$.
\end{lemma}
{\startproof

Let $x^{\boldsymbol\mu}$ and $\tilde{x}^{\boldsymbol\mu}$ denote the coordinate systems on $S$ determined by
$\partial_\mu$ and $\tilde{\partial}_\mu$ respectively, in the manner described in the foregoing lemma.
From this lemma, the range of the coordinates in these two systems are exactly the same,
so that one can define a diffeomorphism $\varphi: S \rightarrow S$ by the condition
$x^{\boldsymbol\mu}(p) = \tilde{x}^{\boldsymbol\mu}(\varphi(p))$.
For each face $f$ in $S$, because the range of values of the coordinates $x^{\boldsymbol\mu}$ and $\tilde{x}^{\boldsymbol\mu}$ over $f$ are the same --- namely the  points in the corresponding face of the canonical 4-simplex in $\R^4$ ---  $\varphi$ maps $f$ back to itself.

Furthermore, the action of $(\varphi \cdot \partial)$ on the coordinate gradients $(\partial_\nu \tilde{x}^{\boldsymbol\mu})$
is given by
\begin{equation}
\label{derivmap}
(\varphi \cdot \partial)_\mu (\partial_\nu \tilde{x}^{\boldsymbol\mu})
:= (\varphi^{-1})^*(\partial_\mu(\varphi^*(\partial_\nu \tilde{x}^{\boldsymbol\mu})))
=  (\varphi^{-1})^*(\partial_\mu \partial_\nu (\varphi^*\tilde{x}^{\boldsymbol\mu}))
=  (\varphi^{-1})^*(\partial_\mu \partial_\nu x^{\boldsymbol\mu}) = 0
\end{equation}
where, in the last step, the fact that $\partial_\mu$ is the coordinate derivative for $x^{\boldsymbol\mu}$ was used.
Equation (\ref{derivmap}) implies that $(\varphi \cdot \partial)_\mu$ is the coordinate derivative for $\tilde{x}^{\boldsymbol\mu}$,
whence $(\varphi \cdot \partial)_\mu = \tilde{\partial}_\mu$.
%
%
%

It remains only to show that $\varphi$ is orientation preserving. This can be seen from the fact that $\varphi$
maps $\frac{\partial}{\partial x^\mathbf{0}}^{[\alpha} \cdots \frac{\partial}{\partial x^\mathbf{3}}^{\delta]}$
to $\frac{\partial}{\partial \tilde{x}^\mathbf{0}}^{[\alpha} \cdots
\frac{\partial}{\partial \tilde{x}^\mathbf{3}}^{\delta]}$. Specifically,
because these two inverse 4-forms are nowhere vanishing, there exists a nowhere vanishing function
$\lambda$, which therefore doesn't change sign, such that
\begin{equation}
\label{invfourforms}
\frac{\partial}{\partial \tilde{x}^\mathbf{0}}^{[\alpha} \cdots \frac{\partial}{\partial \tilde{x}^\mathbf{3}}^{\delta]}
= \lambda
\frac{\partial}{\partial x^\mathbf{0}}^{[\alpha} \cdots \frac{\partial}{\partial x^\mathbf{3}}^{\delta]}.
\end{equation}
To find the sign of $\lambda$, it is
sufficient to find its sign at a single point.  At $p_4$, for each $\boldsymbol\mu$,
by construction, $\frac{\partial}{\partial x^{\boldsymbol\mu}}$ and $\frac{\partial}{\partial \tilde{x}^{\boldsymbol\mu}}$
are both tangent to $\overline{p_4 p_{\boldsymbol\mu}}$ and oriented in the direction of $p_{\boldsymbol\mu}$.
%
%
It follows that the coefficient $\lambda$ in equation (\ref{invfourforms})
is positive at $p_4$, and thus positive throughout $S$. 
Thus the push-forward action of $\varphi$ maps
$\frac{\partial}{\partial x^\mathbf{0}}^{[\alpha} \cdots \frac{\partial}{\partial x^\mathbf{3}}^{\delta]}$
to itself times an everywhere positive function, so that $\varphi$ is orientation preserving.
\finishproof}

\begin{theorem}
$\omega(B_{\mu\nu}^{IJ}(\{\mathcal{B}_{ab}\}, \partial))$ and
$\nu(B_{\mu\nu}^{IJ}(\{\mathcal{B}_{ab}\}, \partial))$ are
independent of the choice of $\partial_\mu$ adapted to $S$.
\end{theorem}
{\startproof
Let $\partial_\mu$ and $\tilde{\partial}_\mu$ be two flat connections adapted to $S$.
Then by the previous lemma, there exists an orientation preserving diffeomorphism
$\varphi: S \rightarrow S$ mapping $\partial_\mu$ to $\tilde{\partial}_\mu$ and such that
$\varphi$ preserves each face of $S$, where `face'
includes in its meaning tetrahedra, triangles, edges, and vertices on the boundary.  In particular, for each $a,b \in \{0,1,2,3,4\}$,
$\varphi$ preserves $\Delta_{ab}(S)$.  Because it also preserves tetrahedron $a$ and $b$ and the fixed orientation of $S$,
it in fact also preserves the orientation of  $\Delta_{ab}(S)$ \cite{engle2011}.
%
%
Using this fact and the diffeomorphism
covariance of the form integral, one has
\begin{equation}
\label{satisfya}
\int_{\Delta_{ab}(S)} \varphi^* B(\{\mathcal{B}_{cd}\}, \tilde{\partial})
= \int_{\Delta_{ab}(S)} B(\{\mathcal{B}_{cd}\}, \tilde{\partial})
= \mathcal{B}_{ab}
\end{equation}
where the definition of  $B_{\mu\nu}(\{\mathcal{B}_{cd}\}, \tilde{\partial})$ was used in the last step.
Furthermore,
\begin{equation}
\label{satisfyb}
\partial_\sigma(\varphi^* B_{\mu\nu}(\{\mathcal{B}_{ab}\}, \tilde{\partial}))
= \varphi^* \circ (\varphi^{-1})^* \partial_\sigma(\varphi^* B_{\mu\nu}(\{\mathcal{B}_{ab}\}, \tilde{\partial}))
= \varphi^* \tilde{\partial}_\sigma B_{\mu\nu}(\{\mathcal{B}_{ab}\}, \tilde{\partial}) = 0
\end{equation}
where again the definition of $B_{\mu\nu}(\{\mathcal{B}_{ab}\}, \tilde{\partial})$ was used in the last step.
Equations (\ref{satisfya}) and (\ref{satisfyb}) then imply
\begin{equation}
B_{\mu\nu}(\{\mathcal{B}_{ab}\}, \partial) = \varphi^* B_{\mu\nu}(\{\mathcal{B}_{ab}\}, \tilde{\partial}).
\end{equation}
Because $\varphi$ is orientation preserving, and both the orientation and Plebanski sector of $B_{\mu\nu}$
are invariant under orientation preserving diffeomorphisms, one has
\begin{displaymath}
\omega(B_{\mu\nu}(\{\mathcal{B}_{ab}\}, \partial)) = \omega(B_{\mu\nu}(\{\mathcal{B}_{ab}\}, \tilde{\partial})) \qquad \text{and} \qquad
\nu(B_{\mu\nu}(\{\mathcal{B}_{ab}\}, \partial)) = \nu(B_{\mu\nu}(\{\mathcal{B}_{ab}\}, \tilde{\partial})),
\end{displaymath}
proving the theorem.
\finishproof}

\section{Four dimensional closure}
\label{closureapp}

The following property is mentioned, for example,
in \cite{cdm1989, cf2008, go2010}.
%
%
\begin{theorem}
For any geometrical 4-simplex $\geosimp$ in $\R^4$,
\begin{equation}
\sum_{t} V_t N_t^I = 0
\end{equation}
where the sum is over tetrahedra, and $V_t$ and $N_t^I$ are the volume and
outward normal to tetrahedron $t$.
\end{theorem}
{\startproof
Define ${}^3 \epsilon^I$ to be the three-form on $\R^4$ with
components $({}^3 \epsilon^I)_{JKL} = \epsilon^I{}_{JKL}$.  Then
%
%
\begin{displaymath}
\dif {}^3 \epsilon^I = 0.
\end{displaymath}
Thus,
\begin{equation}
\label{closurefirst}
0 = \int_\geosimp \dif {}^3\epsilon^I = \sum_t \int_t {}^3\epsilon^I .
\end{equation}
Let ${}^t \epsilon$ denote the volume form for $t$, so that for each $t$,
\begin{displaymath}
\epsilon_{IJKL} = 4 (N_t)_{[I} ({}^t \epsilon)_{JKL]}.
\end{displaymath}
Pulling back $JKL$ to tetrahedron $t$, it follows that
\begin{displaymath}
\underset{t \leftarrow}{{}^3 \epsilon}^I = N_t^I ({}^t \epsilon)
\end{displaymath}
which combined with \ref{closurefirst} yields the result.
\finishproof}

\section{Properties of embeddings and projectors}
\label{projvarapp}

%

Recall $\njm{n}{\rotk}{m}$ denotes the eigenstate of $n \cdot \hat{L}$ in $V_k$
with eigenvalue $m$, and $\njm{n}{\genjpm^-, \genjpm^+, \rotk}{m}$ the eigenstate of $\hat{L}^2$ and $n \cdot \hat{L}$ in $V_{\genjpm^-. \genjpm^+}$
with eigenvalues $k(k+1)$ and $m$,
\begin{lemma}\label{iotalemm}
\dummy
\begin{enumerate}
\item[(a.)]
\begin{equation}
\label{iotaintertwine}
\hat{\rotgen}^i \circ \geniota{}{} =
\geniota{}{} \circ \hat{\rotgen}^i
\end{equation}
\item[(b.)]For each unit $n^i \in \R^3$ and each $k,m$, there exists $\theta \in \frac{\R}{2\pi \Z}$ such that
\begin{equation}
\geniota{}{} \njm{n}{\rotk}{m}
= e^{i \theta} \njm{n}{\genjpm^-, \genjpm^+, \rotk}{m}.
\end{equation}
\item[(c.)]
For any $\mathcal{S} \subseteq \R$,
\begin{equation}
P_{\mathcal{S}} (n \cdot \hat{\rotgen}) \circ \geniota{}{}
= \geniota{}{} \circ P_{\mathcal{S}}(n \cdot \hat{\rotgen}) .
\end{equation}
\end{enumerate}
%
%
\end{lemma}
{\startproof

\noindent\textit{Proof of (a.):}\\
From the intertwining property of $\geniota{}{}$,
\begin{eqnarray}
\rho(e^{t\sut{\tau}^i}, e^{t \sut{\tau}^i}) \geniota{}{} =
\geniota{}{} \rho(e^{t\sut{\tau}^i}).
\end{eqnarray}
Taking $i \frac{\dif }{\dif t}$ of both sides and setting $t = 0$ yields the result.

\noindent\textit{Proof of (b.):}\\
Using part (a.),
\begin{eqnarray*}
\left( n \cdot \hat{\rotgen}\right) \geniota{}{}\njm{n}{\rotk}{m}
&=& \geniota{}{} \left( n \cdot \hat{\rotgen}\right) \njm{n}{\rotk}{m}
= m  \geniota{}{} \njm{n}{\rotk}{m} , \qquad \text{and} \\
 \hat{\rotgen}^2 \geniota{}{}\njm{n}{\rotk}{m}
&=& \geniota{}{} \hat{\rotgen}^2 \njm{n}{\rotk}{m}
= \rotk(\rotk+1)  \geniota{}{} \njm{n}{\rotk}{m} .
\end{eqnarray*}
The result follows.

\noindent\textit{Proof of (c.):}\\
We have for each $m$,
\begin{eqnarray*}
P_\mathcal{S}(n \cdot \hat{\rotgen}) \geniota{}{}
\njm{n}{\rotk}{m} &=&
e^{i\theta} P_\mathcal{S}(n \cdot \hat{\rotgen})
\njm{n}{\genjpm^-, \genjpm^+, \rotk}{m}
= e^{i\theta} \chi_\mathcal{S}(m)
\njm{n}{\genjpm^-, \genjpm^+, \rotk}{m} \\
&=& \chi_\mathcal{S}(m)
\geniota{}{}
\njm{n}{\rotk}{m}
= \geniota{}{} P_\mathcal{S}(n \cdot \hat{\rotgen})
\njm{n}{\rotk}{m}
\end{eqnarray*}
where $\chi_\mathcal{S}(m)$ denotes the characteristic function for $\mathcal{S}$.
%
%
\finishproof}

\begin{lemma}
\label{projgroup_comm}
In any irreducible representation (irrep) of $Spin(4)$,
for any two $(v_-)^i, (v_+)^i \in \R^3$
\begin{equation}
\rho(X^-, X^+) \circ P_{\mathcal{S}}(v_- \cdot \hat{J}^- + v_+ \cdot \hat{J}^+ )
= P_{\mathcal{S}}\left( (X^- \triangleright v_-) \cdot \hat{J}^- + (X^+  \triangleright v_+) \cdot \hat{J}^+ \right)
\rho(X^-, X^+)
\end{equation}
\end{lemma}
{\startproof
Let $\genj^\pm, m^\pm$ be given.
Write $v_\pm^i = \lambda_\pm n_\pm^i$ with $\lambda_\pm \ge 0$ and $n_\pm^i$ unit.
Using that \\
\mbox{$\rho(X^\pm) \njm{n_\pm}{\genj^\pm}{m^\pm} = e^{i\theta^\pm}
\njm{X^\pm \triangleright n_\pm}{\genj^\pm}{m^\pm}$} for some $\theta^\pm$, we have
\begin{eqnarray}
\nonumber
&& \hspace{-1cm} \rho(X^-, X^+) P_{\mathcal{S}}(v_- \cdot \hat{J}^- + v_+ \cdot \hat{J}^+ )
\njm{n_-}{\genj^-}{m^-} \otimes \njm{n_+}{\genj^+}{m^+} \\
\nonumber
&=&  \chi_{\mathcal{S}}(\lambda_- m^- + \lambda_+ m^+ )  \rho(X^-, X^+)
\njm{n_-}{\genj^-}{m^-} \otimes \njm{n_+}{\genj^+}{m^+}\\
\nonumber
&=& e^{i(\theta^- + \theta^+)}\chi_{\mathcal{S}}(\lambda_- m^- + \lambda_+ m^+ )
\njm{X^- \triangleright n_-}{\genj^-}{m^-} \otimes
\njm{X^+ \triangleright n_+}{\genj^+}{m^+}\\
\nonumber
&=& e^{i(\theta^- + \theta^+)}P_{\mathcal{S}}\left((X^- \triangleright v_-) \cdot \hat{J}^-
+ (X^+ \triangleright v_+) \cdot \hat{J}^+\right)
\njm{X^- \triangleright n_-}{\genj^-}{m^-} \otimes
\njm{X^+ \triangleright n_+}{\genj^+}{m^+}\\
\nonumber
&=& P_{\mathcal{S}}\left((X^- \triangleright v_-) \cdot \hat{J}^-
+ (X^+ \triangleright v_+) \cdot \hat{J}^+\right)
\rho(X^-, X^+)
\njm{n_-}{\genj^-}{m^-} \otimes \njm{n_+}{\genj^+}{m^+}
\end{eqnarray}
\finishproof}

\begin{lemma}
Let $\hat{O}_t$ be any one-parameter family of self-adjoint operators
on a Hilbert space $\Hil$. For each $t$, let $\psi_t$ be a normalized
eigenstate of $\hat{O}_t$
such that all $\psi_t$ have the same eigenvalue $\lambda \in \R$.
%
%
Then
\begin{equation}
\langle \psi_t | \left(\frac{\dif}{\dif t} \hat{O}_t\right)  | \psi_t \rangle = 0.
\end{equation}
\end{lemma}
{\startproof
\begin{equation}
\langle \psi_t | \hat{O}_t | \psi_t \rangle = \lambda
\end{equation}
for all $t$.  Taking $\frac{\dif}{\dif t}$ of both sides,
\begin{eqnarray*}
\left(\frac{\dif}{\dif t}\langle \psi_t |\right) \hat{O}_t | \psi_t \rangle
+ \langle \psi_t |\left(\frac{\dif}{\dif t} \hat{O}_t\right) | \psi_t \rangle
+\langle \psi_t | \hat{O}_t \frac{\dif}{\dif t} | \psi_t \rangle &=& 0 \\
\lambda \frac{\dif}{\dif t}\left(\langle \psi_t, \psi_t \rangle\right)
 + \langle \psi_t |\left(\frac{\dif}{\dif t} \hat{O}_t\right) | \psi_t \rangle &=& 0 \\
  \langle \psi_t |\left(\frac{\dif}{\dif t} \hat{O}_t\right) | \psi_t \rangle &=& 0
\end{eqnarray*}
\finishproof}
Applying this to the family of operators $\hat{O}_t = P_{\mathcal{S}}(n_t \cdot \hat{L})$
on $\spinsp_{\genjpm^-, \genjpm^+}$ and the states $\njm{n_t}{\genjpm^-, \genjpm^+, \rotk}{m}$,
and to the family of operators $\hat{O}_t = P_{\mathcal{S}}(n_t \cdot \hat{L})$
on $\spinsp_{k}$ and the states $\njm{n_t}{\rotk}{m}$,
yields the following.
\begin{corollary}
\label{projvarcor}
For any variation $\delta$ of $n$, any $\genjpm^-, \genjpm^+, \rotk$, any $m \in \{-\rotk, -\rotk+1, \dots, \rotk\}$, and any set $\mathcal{S} \subset \R$, one has
\begin{equation}
\njmb{n}{\genjpm^-, \genjpm^+, \rotk}{m} \delta P_\mathcal{S}(n\cdot \hat{L})
\njm{n}{\genjpm^-, \genjpm^+, \rotk}{m}= 0.
\end{equation}
and
\begin{equation}
\njmb{n}{\rotk}{m} \delta P_\mathcal{S}(n\cdot \hat{L})
\njm{n}{\rotk}{m}= 0.
\end{equation}
\end{corollary}

\section{Expression for vertex with projectors on the left}
\label{leftapp}

\begin{lemma}
For each unit $n^i \in \R^3$, $g \in SU(2)$, $k$, and $m$, there exists
$\theta \in \frac{\R}{2\pi\Z}$ such that
\begin{equation}
\label{nstatecov}
\rho(h) \njm{n}{k}{m} =
e^{i\theta} \njm{h \triangleright n}{k}{m}
\end{equation}
\end{lemma}
{\startproof
\begin{displaymath}
\left((h \triangleright n) \cdot \hat{\rotgen}\right) \rho(h) \njm{n}{k}{m}
= \rho(h) (n \cdot \hat{\rotgen}) \njm{n}{k}{m}
= m\rho(h) \njm{n}{k}{m}.
\end{displaymath}
\finishproof}

\begin{lemma}
\label{Jproj_comm}
For any $\mathcal{S} \subseteq \R$, and in any irrep
of $Spin(4)$, and any $v^i \in \R^3$,
\begin{equation}
P_{\mathcal{S}}(v \cdot \hat{\rotgen}) \circ J = J \circ P_{\mathcal{S}}(-v \cdot \hat{\rotgen})
\end{equation}
\end{lemma}
{\startproof
Let $v^i=: \lambda n^i$ with $\lambda \ge 0$ and $n^i$ unit.
Using that $J$ anticommutes with $\hat{\rotgen}^i$, for any $n$ and $\rotk$,
\begin{equation}
\left(n \cdot \hat{\rotgen}\right) J \njm{n}{\genjpm^-, \genjpm^+, \rotk}{m}
= - J \left(n \cdot \hat{\rotgen}\right) \njm{n}{\genjpm^-, \genjpm^+, \rotk}{m}
= -m J \njm{n}{\genjpm^-, \genjpm^+, \rotk}{m}
\end{equation}
whence
\begin{equation}
J \njm{n}{\genjpm^-, \genjpm^+, \rotk}{m} = e^{i \theta_m} \njm{n}{\genjpm^-, \genjpm^+, \rotk}{-m}
\end{equation}
for some $\{\theta_m\} \subset \frac{\R}{2\pi \Z}$,
so that, for all $m$,
\begin{eqnarray*}
P_{\mathcal{S}}(v \cdot \hat{\rotgen}) J \njm{n}{\genjpm^-, \genjpm^+, \rotk}{m}
&=& e^{i\theta_m} P_{\mathcal{S}}(v \cdot \hat{\rotgen}) \njm{n}{\genjpm^-, \genjpm^+, \rotk}{-m} \\
&=& \chi_{\mathcal{S}}(-\lambda m) J  \njm{n}{\genjpm^-, \genjpm^+, \rotk}{m}
= J P_{\mathcal{S}}(-v\cdot\hat{\rotgen}) \njm{n}{\genjpm^-, \genjpm^+, \rotk}{m}.
\end{eqnarray*}
\finishproof}

\begin{theorem}
The vertex amplitude (\ref{propeprl})
can also be written with the projector, appropriately transformed, moved to anywhere in each face factor:
\begin{eqnarray}
\nonumber
A^\prop_v (\{\rotk_{ab}, \psi_{ab}\})
&=& \int_{\rm Spin(4)^5} \prod_a \dif G_a
\prod_{a<b} \epsilon( \simpiota{ab} \psi_{ab},
\rho(G_{ab})  P_{ba}(\{G_{a'b'}\}) 
 \simpiota{ab}\psi_{ba}) \\
\nonumber
&=& \int_{\rm Spin(4)^5} \prod_a \dif G_a
\prod_{a<b} \epsilon( P_{ab}(\{G_{a'b'}\}) \simpiota{ab} \psi_{ab},
\rho(G_{ab})
 \simpiota{ab}\psi_{ba}) \\
\label{leftexp}
&=& \int_{\rm Spin(4)^5} \prod_a \dif G_a
\prod_{a<b} \epsilon( \simpiota{ab}  P_{ab}(\{G_{a'b'}\}) \psi_{ab},
\rho(G_{ab})
 \simpiota{ab}\psi_{ba}) .
\end{eqnarray}
\end{theorem}
{\startproof
One starts from
(\ref{propeprl_herm}), and uses lemma \ref{iotalemm}.c,
lemma \ref{projgroup_comm}, the hermicity of orthogonal projectors, and lemma \ref{Jproj_comm} in succession,
as well as using the fact that if $X^i{}_j$ denotes the adjoint action of $X = SU(2)$ then 
$X^i{}_j \tau^j = X^{-1} \tau^i X$.
%
%
\finishproof}

%

\begin{thebibliography}{10}

\bibitem{feynman1948}
R.~Feynman, ``Space-time approach to non-relativistic quantum mechanics,'' {\em
  Rev. Mod. Phys.}, vol.~20, pp.~367--387, 1948.

\bibitem{feynman1942}
R.~Feynman, {\em The Principle of Least Action in Quantum Mechanics}.
\newblock PhD thesis, Princeton University, 1942.

\bibitem{dirac1930}
P.~A.~M. Dirac, {\em The Principles of Quantum Mechanics}.
\newblock Oxford: Oxford UP, 1st~ed., 1930.

\bibitem{rovelli2004}
C.~Rovelli, {\em Quantum Gravity}.
\newblock Cambridge: Cambridge UP, 2004.

\bibitem{perez2003}
A.~Perez, ``Spin foam models for quantum gravity,'' {\em Class. Quant. Grav.},
  vol.~20, p.~R43, 2003.

\bibitem{rovelli2011}
C.~Rovelli, ``Zakopane lectures on loop gravity,'' {\em Proc. Sci.},
  vol.~QGQGS2011, p.~003, 2011.
\newblock arXiv:1102.3660.

\bibitem{al2004}
A.~Ashtekar and J.~Lewandowski, ``Background independent quantum gravity: A
  status report,'' {\em Class. Quant. Grav.}, vol.~21, p.~R53, 2004.

\bibitem{thiemann2007}
T.~Thiemann, {\em Modern Canonical Quantum General Relativity}.
\newblock Cambridge: Cambridge UP, 2007.

\bibitem{elpr2007}
J.~Engle, E.~Livine, R.~Pereira, and C.~Rovelli, ``{L}{Q}{G} vertex with finite
  {I}mmirzi parameter,'' {\em Nucl. Phys. B}, vol.~799, pp.~136--149, 2008.

\bibitem{epr2007}
J.~Engle, R.~Pereira, and C.~Rovelli, ``The loop-quantum-gravity
  vertex-amplitude,'' {\em Phys. Rev. Lett.}, vol.~99, p.~161301, 2007.

\bibitem{epr2007a}
J.~Engle, R.~Pereira, and C.~Rovelli, ``Flipped spinfoam vertex and loop
  gravity,'' {\em Nucl. Phys. B}, vol.~798, pp.~251--290, 2008.

\bibitem{kkl2009}
W.~Kami\'{n}ski, M.~Kisielowski, and J.~Lewandowski, ``Spin-foams for all loop
  quantum gravity,'' {\em Class. Quant. Grav.}, vol.~27, p.~095006, 2010.

\bibitem{fk2007}
L.~Freidel and K.~Krasnov, ``A new spin foam model for 4d gravity,'' {\em
  Class. Quant. Grav.}, vol.~25, p.~125018, 2008.

\bibitem{bdfgh2009}
J.~Barrett, R.~Dowdall, W.~Fairbairn, H.~Gomes, and F.~Hellmann, ``Asymptotic
  analysis of the {EPRL} four-simplex amplitude,'' {\em J. Math. Phys.},
  vol.~50, p.~112504, 2009.

\bibitem{cf2008}
F.~Conrady and L.~Freidel, ``{On the semiclassical limit of 4d spin foam
  models},'' {\em Phys. Rev.}, vol.~D78, p.~104023, 2008.

\bibitem{mp2011}
E.~Magliaro and C.~Perini, ``{R}egge gravity from spinfoams,'' {\em
  arXiv:1105.0216}, 2011.

\bibitem{hz2011}
M.~Han and M.~Zhang, ``{Asymptotics of Spinfoam Amplitude on Simplicial
  Manifold: Euclidean Theory},'' {\em Class.Quant.Grav.}, vol.~29, p.~165004,
  2012.

\bibitem{engle2011}
J.~Engle, ``{The Plebanski sectors of the EPRL vertex},'' {\em Class. Quant.
  Grav.}, vol.~28, p.~225003, 2011.
\newblock Corrigendum: {\it Class. Quant. Grav.} vol. 30, p. 049501, 2013.

\bibitem{engle2011err}
J.~Engle, ``{Erratum: The Plebanski sectors of the EPRL vertex},'' {\em Class.
  Quant. Grav.}, vol.~30, p.~049501, 2013.
%
%

\bibitem{engle2012}
J.~Engle, ``A spin-foam vertex amplitude with the correct semiclassical
  limit,'' {\em Phys. Lett. B}, vol.~724, pp.~333--337, 2013.

\bibitem{plebanski1977}
J.~Plebanski, ``On the separation of {E}insteinian substructures,'' {\em J.
  Math. Phys.}, vol.~18, pp.~2511--2520, 1977.

\bibitem{df1998}
R.~De~Pietri and L.~Freidel, ``{so(4)} {P}lebanski action and relativistic spin
  foam model,'' {\em Class. Quant. Grav.}, vol.~16, pp.~2187--2196, 1999.

\bibitem{bhnr2004}
E.~Buffenoir, M.~Henneaux, K.~Noui, and P.~Roche, ``Hamiltonian analysis of
  {P}lebanski theory,'' {\em Class. Quant. Grav.}, vol.~21, pp.~5203--5220,
  2004.

\bibitem{holst1995}
S.~Holst, ``Barbero's {H}amiltonian derived from a generalized
  {H}ilbert-{P}alatini action,'' {\em Phys. Rev. D}, vol.~53, pp.~5966--5969,
  1996.

\bibitem{bfh2009}
J.~Barrett, W.~Fairbairn, and F.~Hellmann, ``Quantum gravity asymptotics from
  the {SU}(2) 15j symbol,'' {\em Int. J. Mod. Phys. A}, vol.~25,
  pp.~2897--2916, 2010.

\bibitem{livine2002}
E.~Livine, ``Projected spin networks for lorentz connection: Linking spin foams
  and loop gravity,'' {\em Class. Quant. Grav.}, vol.~19, pp.~5525--5542, 2002.

\bibitem{alexandrov2007}
S.~Alexandrov, ``Spin foam model from canonical quantization,'' {\em Phys. Rev.
  D}, vol.~77, p.~024009, 2008.

\bibitem{baez1999}
J.~C. Baez, ``{An Introduction to spin foam models of quantum gravity and BF
  theory},'' {\em Lect.Notes Phys.}, vol.~543, pp.~25--94, 2000.
\newblock Published in Geometry and Quantum Physics. Edited by H. Gausterer and
  H. Grosse. Springer, Berlin, 2000.

\bibitem{rs2010a}
C.~Rovelli and S.~Speziale, ``{Lorentz covariance of loop quantum gravity},''
  {\em Phys.Rev.}, vol.~D83, p.~104029, 2011.

\bibitem{pereira2007}
R.~Pereira, ``Lorentzian {LQG} vertex amplitude,'' {\em Class. Quant. Grav.},
  vol.~25, p.~085013, 2008.

\bibitem{bdfhp2009}
J.~Barrett, R.~Dowdall, W.~Fairbairn, F.~Hellmann, and R.~Pereira, ``Lorentzian
  spin foam amplitudes: {G}raphical calculus and asymptotics,'' {\em Class.
  Quant. Grav.}, vol.~27, p.~165009, 2010.

\bibitem{bb2001}
J.~Baez and J.~Barrett, ``Integrability for relativistic spin networks,'' {\em
  Class. Quant. Grav.}, vol.~18, pp.~4683--4700, 2001.

\bibitem{ep2008}
J.~Engle and R.~Pereira, ``Regularization and finiteness of the {L}orentzian
  {LQG} vertices,'' {\em Phys. Rev. D}, vol.~79, p.~084034, 2009.

\bibitem{ls2007}
E.~Livine and S.~Speziale, ``A new spinfoam vertex for quantum gravity,'' {\em
  Phys. Rev. D}, vol.~76, p.~084028, 2007.

\bibitem{connelly1993}
A.~Connelly, ``Rigidity,'' in {\em Handbook of Convex Geometry} (P.~Gruber and
  J.~Wills, eds.), North-Holland, 1993.

\bibitem{cf2008a}
F.~Conrady and L.~Freidel, ``{Path integral representation of spin foam models
  of 4d gravity},'' {\em Class.Quant.Grav.}, vol.~25, p.~245010, 2008.

\bibitem{rovelli2005}
C.~Rovelli, ``Graviton propagator from background-independent quantum
  gravity,'' {\em Phys. Rev. Lett.}, vol.~97, p.~151301, 2006.

\bibitem{ar2007}
E.~Alesci and C.~Rovelli, ``{The Complete LQG propagator. I. Difficulties with
  the Barrett-Crane vertex},'' {\em Phys.Rev. D}, vol.~76, p.~104012, 2007.

\bibitem{ar2007a}
E.~Alesci and C.~Rovelli, ``{The Complete LQG propagator. II. Asymptotic
  behavior of the vertex},'' {\em Phys.Rev. D}, vol.~77, p.~044024, 2008.

\bibitem{bmp2009b}
E.~Bianchi, E.~Magliaro, and C.~Perini, ``{LQG propagator from the new spin
  foams},'' {\em Nucl.Phys. B}, vol.~822, pp.~245--269, 2009.

\bibitem{bd2011}
E.~Bianchi and Y.~Ding, ``{Lorentzian spinfoam propagator},'' {\em Phys.Rev.},
  vol.~D86, p.~104040, 2012.

\bibitem{rv2008}
C.~Rovelli and F.~Vidotto, ``Stepping out of homogeneity in loop quantum
  cosmology,'' {\em Class. Quant. Grav.}, vol.~25, p.~225024, 2008.

\bibitem{brv2010}
E.~Bianchi, C.~Rovelli, and F.~Vidotto, ``Towards spinfoam cosmology,'' {\em
  Phys. Rev. D}, vol.~82, p.~084035, 2010.

\bibitem{vidotto2011}
F.~Vidotto, ``{Many-nodes/many-links spinfoam: The homogeneous and isotropic
  case},'' {\em Class.Quant.Grav.}, vol.~28, p.~245005, 2011.

\bibitem{wald1984}
R.~M. Wald, {\em General Relativity}.
\newblock Chicago: Chicago University Press, 1984.

\bibitem{cdm1989}
M.~Caselle, A.~D'Adda, and L.~Magnea, ``{R}egge calculus as a local theory of
  the {P}oincar\'e group,'' {\em Phys. Lett.}, vol.~B232, p.~457, 1989.

\bibitem{go2010}
S.~Gielen and D.~Oriti, ``{Classical general relativity as BF-Plebanski theory
  with linear constraints},'' {\em Class. Quant. Grav.}, vol.~27, p.~185017,
  2010.

\end{thebibliography}
%
%
%
%

\end{document}